\documentclass[prd,epsf,amsmath,superscriptaddress,citeautoscript,
showpacs,floatfix,reprint, preprintnumbers]{revtex4-2}
\usepackage{bm}
\usepackage{amssymb}
\usepackage{graphicx}
\usepackage{amsmath, amsthm, amssymb, graphicx}
\usepackage[usenames]{color}
\graphicspath{{figures/}} 
\usepackage{ulem}

\usepackage{amsmath,amsthm,amsfonts,amscd,amssymb}
\usepackage{eucal} 	 	
\usepackage{verbatim}      	
\usepackage{makeidx}       	
\usepackage{braket}

\usepackage{natbib}
\usepackage{bibentry}
\usepackage[hidelinks]{hyperref}
\newcommand*{\boldone}{\text{\usefont{U}{bbold}{m}{n}1}}

\usepackage{mathrsfs}
\usepackage{graphicx}
\usepackage{dcolumn}
\usepackage{bm}

\begin{document}

	\title{Testing Quantum Gravity using Pulsed Optomechanical Systems}

\author{Jordan Wilson-Gerow}
    \affiliation{Walter Burke Institute for Theoretical Physics, California Institute of Technology, Pasadena, CA 91125}
	\author{
	Yanbei Chen}
	\affiliation{Theoretical Astrophysics, Cahill,
		California Institute of Technology, 1200 E. California
Boulevard,
		MC 350-17, Pasadena CA 91125, USA}		
	\author{ P.C.E. Stamp}
	\affiliation{Department of Physics and Astronomy, University
		of British Columbia, 6224 Agricultural Rd., Vancouver,
B.C., Canada
		V6T 1Z1}
	\affiliation{Theoretical Astrophysics, Cahill,
		California Institute of Technology, 1200 E. California
Boulevard,
		MC 350-17, Pasadena CA 91125, USA}
	\affiliation{Pacific
		Institute of Theoretical Physics, University of British
Columbia,
		6224 Agricultural Rd., Vancouver, B.C., Canada V6T 1Z1}

\vspace{5mm}

	\begin{abstract}
		
An interesting idea, dating back to Feynman \cite{feynman57},
argues that quantum mechanics may break down for large masses if one entertains the possibility that gravity can be ``classical'', thereby leading to predictions different from
conventional low-energy quantum gravity. Despite the technical difficulty in testing such deviations, a large number of experimental proposals have been put forward due to the high level of fundamental interest.
Here, we consider the Schr\"odinger-Newton (SN) theory and the Correlated Worldline (CWL) theory, and show that they can be distinguished from conventional quantum mechanics, as well as each other, by performing pulsed optomechanics experiments. For CWL specifically we develop a framework resembling the commonly used ``Heisenberg-picture'' treatment of coupled oscillators, allowing one to perform simple calculations for such systems without delving into the deeper path-integral formalism. We find that discriminating between the theories will be very difficult until experimental control over low frequency quantum optomechanical systems is pushed much further. However, the predicted departures of SN and CWL from quantum mechanics occur at the same scale, so both alternative models could in principle be probed by a single experiment.

	\end{abstract}
	\preprint{CALT-TH 2023-042}
	\maketitle

\section{Introduction}
 \label{sec:1-intro}

\subsection{Motivations}

Ever since quantum mechanics (QM) was discovered, reservations
have been posed about its applicability at macroscopic scales. In
recent decades QM has been tested in experiments on
superconducting SQUID devices \cite{chiorescu03}, which confirmed
the QM predictions made by Leggett et al. \cite{ajl87}; similar tests were also done in magnetic systems \cite{taka11}. A key
question - of just how macroscopic were the superpositions
involved in the SQUID tests - is still being debated
\cite{arndt14,whaley,ajl16}.

However, such experiments do not test the most widely discussed mechanism for a breakdown of QM, that coming from gravitation \cite{feynman57,feynmanGR,kibble1,penrose96}. To test this requires ``mass superpositions", in which a sufficiently massive object is quantum delocalized. Such experiments are very difficult - the largest `2-slit' mass superpositions achieved so far involve molecules with mass $\sim 34,000$ amu \cite{arndt14}.

At issue here are two of the biggest questions in physics, viz.,
(a) how can one reconcile QM with General Relativity (GR), which governs physics at large scales; and (b) how to reconcile quantum superpositions with the everyday macroscopic world of definite classical states (often called the ``measurement problem"). 

The claim is that these problems may be resolvable in a theory wherein gravity causes a breakdown of QM. Candidates for this theory \cite{carney19} include:

(i) \textit{Collapse Models:} A phenomenological non-relativistic approach in which a
universal noise field, arising from gravity, causes collapse of the Schr\"odinger wave-function \cite{diosi,bahrami14,bassi13}. Recent experiments seem to have ruled out this ``CSL collapse" idea
\cite{arnquist22,donadi21}, both for general collapse scenarios, and for the `gravitational collapse' version.

(ii) \textit{Schr\"odinger-Newton}: A semiclassical approach, again non-relativistic, in which gravity modifies the Schr\"odinger equation to a
``Schr\"odinger-Newton" (SN) equation for the wave-function dynamics \cite{penrose96,yanbei13a,yanbei17}, leading to a dynamics different from standard QM for this wave-function. Approaches like this have been criticized because they allow superluminal signal propagation \cite{polchinski91,gisin91}, and for other reasons \cite{BLHu14}.  

A careful examination of these criticisms reveals that such superluminal signal propagations all arise from the prescription that once a measurement is made locally, a system's wavefunction simultaneously collapse across the entire constant-time slice of space-time. Superluminal signal propagation can then be circumvented if the wavefunction used to generate gravity at each point is instead taken to be the ``local conditional state'' that is only sensitive to measurement results within the past light cone, leading to a ``causal conditional'' formulation~\cite{yanbei23}, which we shall adopt in this paper.  We do caution that while this formulation avoids superluminal signal propagation, it has not been shown to be generalizable to full general relativity with general covariance.


(iii) \textit{Correlated Worldlines}: A covariant relativistic field-theoretic approach in which the fundamental objects are the path integral ``histories" of the
matter and metric quantum fields. Although these fields are
quantized, their histories are coupled by gravity (also
quantized), and this modifies their dynamics and causes a
breakdown of the superposition principle \cite{BCS18,CWL3}. For
large masses this leads to a ``path-bunched" dynamics which
eventually becomes classical. The form of this  ``Correlated
Worldline'' (CWL) theory is actually fixed by very general
considerations, with no adjustable parameters, and it turns out to
satisfy all the relevant Ward identities \cite{CWL2}, has a
consistent classical limit (Einstein's theory), and consistent
expansions in both Newton's constant $G_N$ and $\hbar$.

Each of these theoretical approaches helps to give a framework for
experiments to work in - in principle they give predictions
different from QM and from each other. Such theoretical frameworks
are very important for any experimental tests - it is hard to test
a theory unless one can compare specific predictions with those of
competitors.

What then is the best way to test theories of this kind? Feynman's
original thought experiment \cite{feynman57,feynmanGR} involved a
2-path system, and there have been several analyses of 2-path
experiments since then, both phenomenological
\cite{kibble1,page81}, and microscopic \cite{CWL3,2-path}.

However a genuine 2-path interference experiment for a massive
object is forbiddingly difficult, simply because the de Broglie
wavelength is far too short (for an object having the Planck mass
$M_p$, moving at $\sim 1$m/sec, the de Broglie wavelength $\lambda
\sim 5 \times 10^{-27}$m). Instead the interest of the community
has focused primarily on optomechanical experiments, and there is
good reason for this - systems such as LIGO have shown how
fantastically sensitive an optomechanical setup can be, in spite
of a large variety of noise and decoherence sources
\cite{adhikari14}. In this paper we investigate a class of tests
that may be done with optomechanical experiments, and provide
detailed predictions for the Schr\"odinger-Newton and CWL theories,
and for QM.

\subsection{Optomechanical Experiments}
 \label{sec:1-optoM}

The sensitivity of some optomechanical systems is such that if QM is obeyed at the mass scale of the LIGO mirrors (ie., 40 kg), then one may at some point expect to see clear evidence for the macroscopic quantum behaviour of these mirrors \cite{KST02}. Such an observation would constitute an increase in the mass scale at which QM has been verified by a factor of $7 \times 10^{23}$, and would thus certainly have extremely important consequences for future physics.

However, as we will argue here, it will not in fact be so easy to eliminate competitor theories to QM; and in optomechanics
experiments it will require advancing the frontier of present experimental capabilities. This is because the competing theories only show sharp differences from QM under certain circumstances, i.e. for non-linear measurements and/or non-Gaussian states. Thus, the Gaussian-shaped quantum delocalization most commonly achieved in optomechanical systems will look almost identical whether the physics is fundamentally described by QM, Schr\"odinger-Newton, or CWL theory.

So far ideas for optomechanical tests of QM at the macroscopic
scale mostly involve, in one way or another, the probing of the motion of macroscopic mechanical oscillators. The use of an optomechanical set-up to test low-energy quantum gravity was apparently first suggested 20 yrs ago \cite{marshall03,bouwm08}, within the context of Penrose's idea \cite{penrose96} that self-gravitational effects could dephase state superpositions of a massive body; this idea was later discussed using a Schr\"odinger-Newton equation for the
wave-function dynamics \cite{penroseSN,SN}, as well as in the
related semiclassical gravity approach \cite{semicl}.

There have been many more recent proposals for optomechanical tests of QM. Perhaps the best known example is the observation of suspended test masses (the LIGO system being a classic example of this). At first glance these experiments \cite{aggarwal20,mcCuller20} indicate that the test mirrors are behaving quantum mechanically - however, as noted above, we will show in this paper that the issue is more subtle. Other proposals include the measurement of nanosphere dynamics \cite{AspNano21} and membranes \cite{jayich08}. General reviews of this field also exist \cite{aspRMP14,yanbei13b}.

Most of these optomechanical designs involve quantifiable mass displacements between the elements of two or more quantum
states in a state superposition, ie., they involve mass superpositions. In what follows we will develop a theoretical framework which allows us to analyze a variety of different experiments designed to look for departures from QM in such mass superpositions. One important piece of this work is the quantitative discussion of the effects of thermal noise and other environmental effects. The other is a new convenient framework for studying oscillator systems within the CWL theory. CWL is based upon an infinite number of coupled systems in a path-integral formalism, and may be perceived as technically foreboding. To remove this obstacle we have developed a simple framework, specifically for studying oscillators, which resembles Heisenberg picture oscillator creation/annihilator operators calculus.

Provided one ignores the coupling of an optomechanical system to its environment, optomechanical experiments are in principle straightforward to analyze, since  we are essentially just dealing with a set of coupled oscillators (of which one or more can be considered to be macroscopic). 

Most of this paper will be focused on the example of a fairly simple optomechanical system, which nevertheless captures all of the essential features we require. This will be a single optical cavity, fed by an external driving laser, in which one of the two mirrors is mobile. The cavity optical field then couples to the mechanical motion of the mirror - both motions are
quantized, and we are interested in the way in which the dynamics of a macroscopic mirror is predicted to be affected by gravity, in the case of 3 different theories, viz., (i) traditional QM, (ii) CWL theory, and (iii) Schr\"odinger-Newton theory. 

We also consider a specific experimental design, which has been widely discussed in the present context, this being a pulsed optomechanics setup \cite{aspelPO}, in which the external laser is pulsed. We propose a specific protocol based on a series of pulses. This test is designed such that one expects a certain outcome to occur with probability \textit{zero} in QM, but with a non-zero probability in both SN and CWL theory.

\subsection{Organization of Paper}
 \label{sec:1-Org}

Our paper is organized as follows: we begin in section 2 with a
presentation of the optomechanical set-up, designed for readers who do not work in quantum optics.  The role of an dissipative coupling to the environment also turns out to be important, particularly for the CWL theory where it facilitates ``path bunching" \cite{CWL3,stamp15}.

In section 3 we briefly describe the SN approach to low-energy quantum gravity; the description is fairly brief since this has been reviewed elsewhere, and previous discussions of tests of QM in SN theory already exist. In section 4 we recall the salient details of the CWL theory, and then in section 5 we then give a detailed discussion of the behaviour of a simple harmonic oscillator in CWL theory. In section 6 we then move on to the description of a cavity optomechanics experiment in CWL theory. 

All of this prepares the ground for a discussion of experiments using the three theories. Section 7 focuses on the specific case of a pulsed optomechanics experiment. One can compare the ``ground states" of the system for the three theories (QM, CWL, and SN), insofar as these states are properly defined. Finally, section 8 gives a general discussion of our results, and our conclusions. 

We find that for typical values of experimental parameters currently prevailing, the difference in predictions of the three theories for optomechanical experiments are very small - however, we see that they can be calculated, with no adjustable parameters. We thus conclude that to test these theories it will require significant advances in the  control of quantum states of low frequency massive oscillators (eg. torsional pendulums).

\section{Conventional Cavity Optomechanics}
 \label{sec:2-optoM}

Before looking at how an optomechanical system will behave in theories where QM breaks down, we first consider how it behaves in conventional theory. By this we mean a theory in which QM is obeyed at low energies, and wherein low-energy quantum gravity is described by a path integral in which one integrates out the quantized metric field $g^{\mu\nu}(x)$ according to the usual rules of quantum field theory (QFT).

\subsection{QM Description of Optomechanical Experiments}
 \label{sec:2.A optoM-QM}

\begin{figure}[ht]
\centering
\includegraphics[scale=0.35]{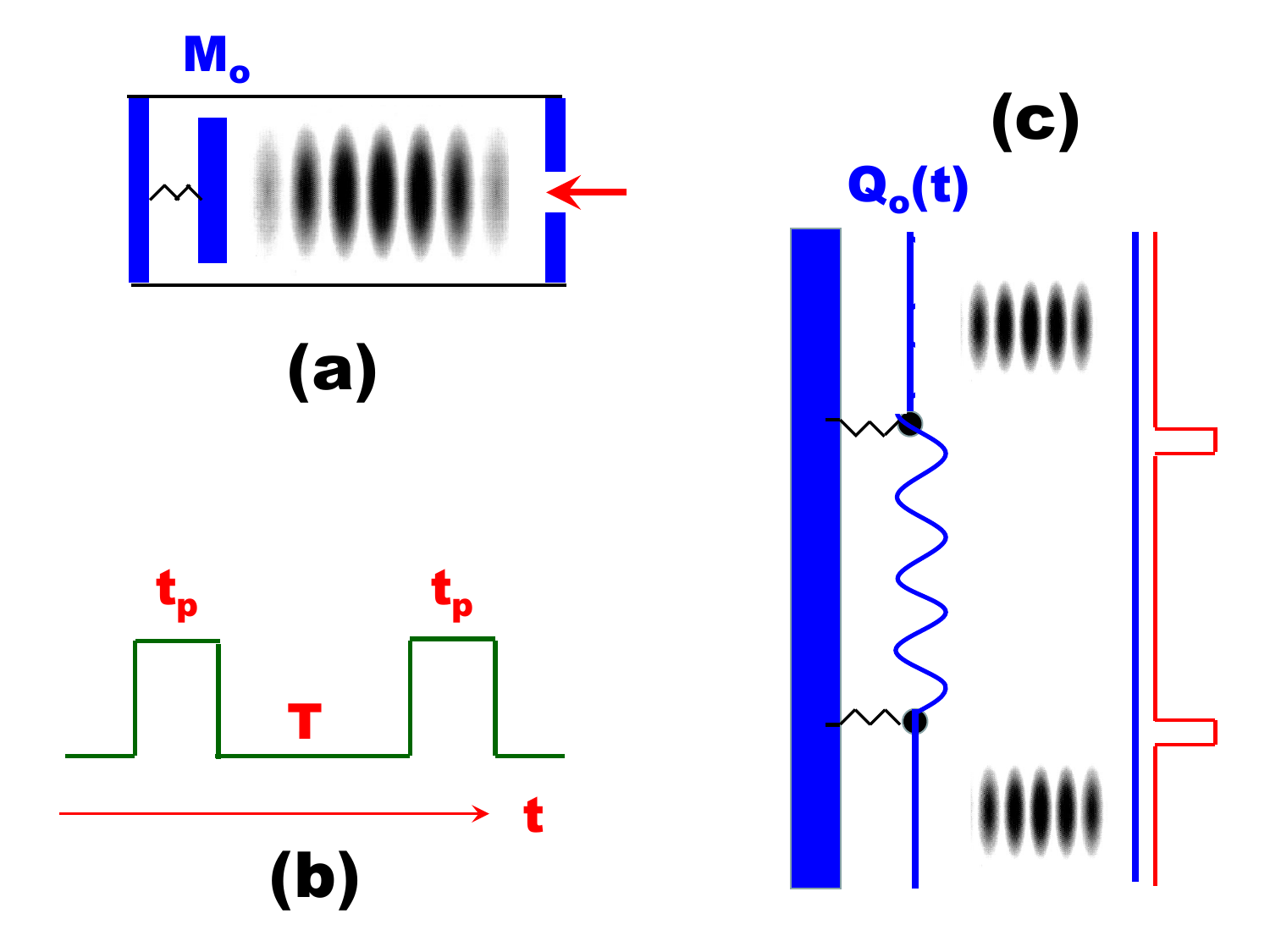}
\vspace{-7mm}
\caption{The optomechanical system considered in the text. In (a) we show the cavity schematically; the laser field is injected from the right, and the oscillating mirror of mass $M_o$ is on the left side of the cavity. In (b) the pulse sequence protocol is shown as a function of time. In (c) we show a spacetime diagram for a pulsed optomechanics experiment, with time along the vertical axis; the initial state (at bottom) and the final state (at top) have all the energy in the cavity mode, with a stationary mirror; the intermediate state has an oscillating mirror. The mirror coordinate is $Q_o(t)$.    }
 \label{fig:cavity}
\end{figure}

There is a large variety of different experimental optomechanical systems \cite{scully}, and one can certainly treat each of these individually. However the key physics is summarized in the setup shown in Fig. \ref{fig:cavity}(a), where one assumes a cavity optical mode with resonant frequency $\omega_{\textrm{cav}}$ to be excited by an external drive laser, and wherein one of the cavity mirrors, of mass $M$, is mobile and behaves as a harmonic oscillator with frequency $\omega_m$; the other mirror is fixed rigidly.

If both mirrors were dynamical, or there was another mechanical oscillator nearby, conventional low-energy quantum gravity predicts  graviton-mediated correlations between the massive oscillators' motion. Genuine experimental proposals based on variants of this set-up have been put forth to measure these correlations, and thereby test conventional quantum gravity at low-energies \cite{et17:23}. In this paper we will focus only on systems with just one moveable mass, so that conventional gravitational effects are negligible.

If there are departures from QM to be seen in such an optomechanical system, it is most natural to assume they will be shown in the dynamics of the massive moveable mirror. This assumption simply reflects the belief that QM has been adequately tested on microscopic systems, and on excitations of the EM field $A^{\mu}(x)$. If one is interested in departures from QM caused by gravity, we note that interactions between $A^{\mu}(x)$ and the metric field $g^{\mu\nu}(x)$ are here completely negligible.

To be specific we assume the moveable mirror to be in the form of a cylindrical plate, of radius $R_o$, thickness $L_o$, and density $\rho_o$.  The internal structure can be amorphous or crystalline, and as we note below, some quantitative differences between the two can show up in experiments.

A detailed description of the mirror and its interaction with the quantized EM field is actually quite complicated. In particular one needs to account for:

(i) the structure of the mirror surface. This may be very complex, with multiple layers having different dielectric properties. At length scales $\gg a_o$, the interatomic spacing, one may model these using a position dependent dielectric function $\epsilon ({\bf r}, \omega)$. However the microscopic structure of the mirror, both in the surface layers and in the bulk, is also relevant, because dynamic defects (often modeled as `two level systems') will interact with both mechanical and EM modes.

(ii) both the fundamental mechanical mirror coordinate and the other higher mechanical modes interact dissipatively with the mirror supports, whatever these may be - this interaction involves microscopic defects of various kinds.

In a simplified treatment these details are ignored, and one uses a model in which the pressure on the mirror from the cavity optical mode gives a simple linear bilinear coupling $\lambda Q A^{\dagger}A$ between the mirror centre of mass coordinate $Q$ and the cavity photon operators $A^{\dagger}, A$, with a  coupling $\lambda = -\partial_{Q}\omega_{\textrm{cav}}(Q)|_{Q=0}$. The cavity photons also couple to an external driving laser, whose `classical' amplitude inside the cavity is written as
\begin{equation}
     \alpha_{\textrm{in}}(t)=e^{-i\omega_{\textrm{L}}t}\omega_{\textrm{cav}}\frac{1}{\sqrt{\kappa}}\alpha(t)
 \label{driveL}
 \end{equation}
where $\alpha(t)$ is a slowly varying (relative to the main frequency $\omega_{\textrm{L}}$) envelope pulse profile; this amplitude gives the mean number of photons sustained in the cavity mode by the laser.


Our system is then that illustrated in Fig. \ref{fig:cavity}(a). The Hamiltonian for this system is written in the form of
a pair of coupled oscillators with a driving force coupled to the optical cavity oscillator, viz.,
\begin{eqnarray}
 \label{eq:optomechhamiltonian}
 H &=& \left[ \frac{P^{2}}{2M}+\frac{M\omega_{m}^{2}}{2}Q^{2} \right] + \omega_{\textrm{cav}}(A^{\dagger}A + \tfrac{1}{2}) \nonumber \\
&& \qquad  -\lambda Q A^{\dagger}A \;-\; \sqrt{\kappa}(A^{\dagger}\alpha_{\textrm{in}}+A\alpha^{*}_{\textrm{in}})
 \end{eqnarray}
where  $\sqrt{\kappa}$ is a phenomenological coupling constant between the driving laser and the cavity mode in question.

We also introduce operators $a,a^{\dagger}$, with $a = A-\alpha$, which describe photon fluctuations about the mean photon amplitude in the cavity. In the same way we define ``phonon" operators $b,b^{\dagger}$ for the mirror coordinate $Q$.

We then have
\begin{equation}
    H=\omega_{m}b^{\dagger}b+\omega_{\textrm{cav}}a^{\dagger}a-g(t)(b+b^{\dagger})(ae^{i\omega_{\textrm{L}}t}+a^{\dagger}e^{-i\omega_{\textrm{L}}t})
    \end{equation}
plus irrelevant constants. We have here defined an optomechanical coupling strength as
\begin{equation}
g(t)=\lambda z_o \alpha(t)
 \label{g-t}
\end{equation}
where $z_o$ is the zero point amplitude for the moveable mirror. Notice that $g(t)$ essentially measures the strength of the indirect interaction between the incoming laser field and the mechanical oscillator, simply because the cavity photon density is entirely dominated by this field, with only small fluctuations around it. Thus, by switching on and off the laser field, we can to first approximation switch on and off the optomechanical coupling, neglecting only the correction due to small photon fluctuations.

Finally, let us pass to the interaction picture, with $b\rightarrow e^{-i\omega_{m}t}b,\,\,a\rightarrow e^{-i\omega_{\textrm{cav}}t}a$, and define the laser detuning frequency $\Delta=\omega_{\textrm{L}}-\omega_{\textrm{cav}}$. Our quantum optomechanical Hamiltonian is now given by the simple form
\begin{equation}
 \label{eq:Hintoptomech}
    H_{int}=-g(t)(be^{-i\omega_{m}t}+b^{\dagger}e^{i\omega_{m}t})(ae^{i\Delta t}+a^{\dagger}e^{-i\Delta t})
\end{equation}
so that we have isolated out the effect of the interaction between the cavity fluctuation photons and the mechanical quanta, with the coupling strength $g(t)$ being proportional to the driving laser field amplitude.

Consider now the effect of the driving laser on the system.  We note from eqn. (\ref{eq:Hintoptomech}) that in the `red-detuned' case when $\Delta = - \omega_m$, the effect of the driving laser field in the combined space of the optical/mechanical oscillations is simply to `rotate' the basis - ie., we can swap the energy back and forth between the 2 oscillation modes. Let's define the phase factor
\begin{equation}
    \phi(t) \;=\; \int^{t}_{0} d\tau g(\tau)
 \label{phi-t}
\end{equation}
which gives the integrated effects of the coupling over time. Again in the resonant case, we see that when $\phi(t) = (n+\tfrac{1}{2}) \pi/2$ we precisely swap the states of the mechanical and optical oscillators, whereas when $\phi(t) = n\pi$, we return the system to its $t=0$ state; here $n$ is an integer.

This suggests applying a pulse sequence exactly analogous to that in spin echo experiments. We start at $t=0$ with the mechanical oscillator in its ground state and the cavity in some excited state, and then apply a $\pi/2$ pulse, effectively loading the cavity state into the mechanical oscillator. We then let the system evolve - and later we apply a second $\pi/2$ pulse. If the system has evolved coherently, without decoherence, then we should recover the initial $t=0$ state.

Formally, we choose a pulse sequence such that $g(t)$ has the form
\begin{eqnarray}
 g(t)&=&\frac{(2n+1)\pi}{2t_{p}}\bigg[\textrm{rect}\left(\frac{t-\tfrac{1}{2}t_{p}}{t_{p}}\right) \nonumber\\
 && \qquad\qquad + \;  \textrm{rect}\left(\frac{t-\tfrac{1}{2}t_{p}-(T+t_{p})}{t_{p}}\right)\bigg]
 \label{pulseS}
\end{eqnarray}
which describes a rectangular pulse of duration $t_{p}$, followed by a free evolution for time $T$, and then another rectangular pulse of duration $t_{p}$. The amplitude is chosen such that $\phi(t)=(n+\tfrac{1}{2})\pi$ between the pulses and $\phi(t)=(2n+1)\pi$ after both pulses, for some positive integer $n$. This pulse sequence is shown as a function of time $t$ in Fig. \ref{fig:cavity}(b).

It is also useful to plot the time evolution of the system, during this pulse sequence, on a spacetime diagram. This is shown in Fig. \ref{fig:cavity}(c). Physically what happens is that in the initial (unpulsed) state the mirror is stationary; all of the energy is in the cavity mode. The first pulse is then applied, and the mirror now oscillates, having acquired the energy from the cavity mode. The second pulse then returns the energy to the cavity mode, and the mirror is again stationary.

As we will see, the interest of this sort of protocol for testing QM lies in the fact that corrections to QM caused by gravity can alter the mirror dynamics, and prevent the system from returning precisely to its initial state, after the sequence of pulses just described. Depending on what sort of theory we are dealing with, this may or may not look like a decoherence effect.  By construction, such a pulse protocol will leave \textit{zero} excitations in the mirror degree of freedom if QM is the correct description of the dynamics whereas (as we will see in later sections) both SN and CWL predict a \textit{nonzero} occupation number, so this experiment could in principle distinguish between QM, SN, and CWL theory. 

Schemes like this, where phonon/photon numbers are measured, are sometimes referred to as `non-linear measurement' protocols. Moreover the intermediate state is a Fock state excitation which is an example of a `non-Gaussian state'.  It is not obvious a priori why linear measurements and Gaussian states are not sufficient for distinguishing the three theories, and this will be explained in detail in a later section. Given that we have not yet introduced the technical description of SN and CWL we must leave it for the moment as a statement of fact that, at least for CWL theory, \textit{linear measurements on Gaussian states are insensitive to departures from QM}.

\subsection{Dissipation and Decoherence}
 \label{sec:2.B dissEnv}

Any complete description of the optomechanical system also has to include its coupling to background environmental degrees of freedom, which lead to both dissipation and decoherence effects. If the optomechanical system is in thermal equilibrium with the environment, then this environment also sets the effective temperature of the system. Since any tests of QM or alternative theories to QM may be sensitive to temperature as well as to dissipation and decoherence effects, it is important to quantify them. 

Here we just describe the key environmental couplings and how they are parameterized - their effects are discussed below.

The optomechanical coordinate $Q$ and its conjugate momentum $P$ couple to two kinds of environment, as follows:

(i) There is a coupling to delocalized modes, modelled themselves by oscillator coordinates $(q_k,p_k)$; we write the Hamiltonian as $H_{\mbox{\scriptsize eff}} = H_o (P,Q ) + H^{osc}$, where $H_o (P,Q)$ is just the mechanical oscillator Hamiltonian in eqn. (\ref{eq:optomechhamiltonian}), and where the `oscillator bath' part of the Hamiltonian is \cite{feynmanV63,CalL83,ajl84}
\begin{eqnarray}
H^{osc} \; &=& \;   \sum_{k=1}^{N_o}
\bigg[ F_k(P,Q)x_k +G_k(P,Q)p_k \bigg] \nonumber \\
\;\;  && \qquad + \; {1 \over 2} \sum_{k=1}^{N} \left(  {p_k^2 \over m_k} + m_k \omega_k^2 x_k^2 \right)   
\label{2.3}
\end{eqnarray}
The couplings $F_k(P,Q)$ and $G_k(P,Q)$ are
$\sim O(N_o^{-1/2})$, so that in the "thermodynamic limit" where the number of bath oscillators $N_o \gg 1$, appropriate to a macroscopic environment of delocalised oscillators, these couplings are small. 

(ii) There is a coupling to localized modes (primarily defects in the mirror coatings), which are commonly modelled as a set of 2-level systems, with a `spin bath' environmental term \cite{PS00,gaita}:
\begin{eqnarray}
H^{SB} \; &=& \;  
{\cal F}(P,Q; \{ \mbox{\boldmath $\sigma$}_{\mu} \} ) \nonumber \\ && \qquad \;+\;
\sum_{\mu}^{N_s} {\bf h}_{\mu} \cdot \mbox{\boldmath $\sigma$}_{\mu} \;+\;
\sum_{\mu,\mu'}^{N_s} V_{\mu\mu'}^{\alpha \beta} \sigma_{\mu}^{\alpha}
\sigma_{\mu'}^{\beta}
 \label{H-SB}
\end{eqnarray}

Here the Pauli spins $\{ \mbox{\boldmath $\sigma$}_{\mu} \}$ are the environmental variables, with $\mu = 1,2,...N_s$, and the couplings  ${\cal F}(P,Q; \{ \mbox{\boldmath $\sigma$}_{\mu} \} )$ and the local fields ${\bf h}_{\mu}$ may now be independent of $N_s$, and are not necessarily weak (unlike the couplings in the oscillator bath model). These couplings are in general functions of the entire set of bath spins. 

The interactions $V_{\mu\mu'}^{\alpha \beta} \sigma_{\mu}^{\alpha}\sigma_{\mu'}^{\beta}$ between the spin bath modes can be weak or strong depending on what these modes are (the dipolar interactions between defects in the mirror coating can be strain-mediated or electric dipole interaction mediated, with nearest-neighbour interactions of strengths up to hundreds of Kelvin). 

For any well-designed mirror one typically assumes that these environmental effects are weak and that they can be parameterized by an effective $Q$-factor for the mirror, with $Q \gg 1$. This amounts to assuming an oscillator bath model for the environment, with the additional assumption that this bath is in equilibrium at some temperature $T$. If the coupling of the spin bath modes to the mirror is weak, and if the inter-defect coupling is weak, then one can map the spin bath to an oscillator bath \cite{PS00}, justifying this step. In our view this assumption needs further investigation, but we will adopt it here. 

Under these (quite stringent) circumstances one can describe the classical behaviour of the mirror as a classical damped oscillator. The quantum dynamics is described by a quantum oscillator coupled to a quantized oscillator bath. Both of these models have been studied for conventional quantum systems in great depth \cite{weiss99}. Later in this paper we will discuss what happens in the CWL and SN models.

\section{Schr\"odinger-Newton Theory}
 \label{sec:3-optoM-SchN}

The Schr\"odinger-Newton (SN) theory is a non-relativistic modification of the Schr\"odinger equation in which QM breaks down because of a ``self-interaction" mediated by the Newtonian gravitational interaction. It us thus a theory in which the matter particles are quantized (non-relativistically) and the Newtonian gravitational field is treated classically. 

From this point of view one can perhaps think of SN theory as a non-relativistic variant of a semiclassical gravity theory (we recall that semiclassical gravity is defined by the field equation $G_{\mu\nu}(x) = \kappa \langle T_{\mu\nu} (\psi(x)) \rangle$, where $\psi(x)$ is the matter wave-function, and $\kappa = 8 \pi G_N/c^4$ in MKS units).  SN theory has been extensively reviewed \cite{bahrami14,bassi13}. 

For our purposes we wish to consider SN theory for an extended body. The Hamiltonian of the center of mass of an extended object with mass $M$ is given in SN theory by
\begin{equation}
H = \frac{\mathbf{p}^2}{2M} + V_{\rm NG} +  V_{G}(\mathbf{x}, \psi)
\end{equation}
where $V_{\rm NG}$ is the non-gravitational part of the potential energy, and $V_G$ is the gravitational potential, which depends on the objects centre of mass wave-function $\psi({\bf x})$.  In the regime when the quantum uncertainty in the centre of mass motion is much less than the zero-point uncertainty of nuclei near their equilibrium positions, this can be simplified into
\begin{equation}
V_G =\frac{M\omega_{\rm SN}^2}{2} (\mathbf{x} -\langle \mathbf{x} \rangle)^2
\end{equation}
where $\Omega_{\rm SN}$ is the Schr\"odinger-Newton frequency, and $\langle\mathbf{ x} \rangle$ the expectation value of the position in the quantum state $\psi$.   

The structure of semiclassical gravity becomes non-unique when measurements are made.  First of all, it is clear that {\it after} a projective measurement is made at $t_0$, the quantum state $\psi$ should be projected to the eigenstate $\psi'$ consistent with the result of measurement -- as already noted by Page and Geilker \cite{page81}.  However, if we use the collapsed state function $\psi'$ for times greater than $t_0$, this is not frame independent, and will allow superluminal signaling, as Polchinski noted somewhat later \cite{polchinski91}.  In the Causal Conditional formulation of semiclassical gravity \cite{yanbei23}, for measurements made at space-time event $(t_0,\mathbf{x}_0)$, one will incorporate its result in the future light cone of this event.

For experiments whose light crossing time is much shorter than the time scale of the experiment,  the Causal Conditional formulation has a negligible difference from a theory in which one instantly collapses the quantum state.  Let us consider a test mass with mass $M$ monitored by light, with phase modulation of light due to the test masses detected via homodyne detection.   The Stochastic Master Equation for this is then given by \cite{yanbei23}
\begin{equation}
d\hat \rho = -\frac{i}{\hbar}[\hat H,\hat\rho] dt -\frac{\alpha^2}{4}[\hat x,[\hat x,\hat\rho]] dt
-\frac{i\alpha}{\sqrt{2} }[\hat x ,\hat\rho] dW
\end{equation}
where
\begin{equation}
\label{SN_H}
\hat H = \frac{\hat p^2}{2m } +\frac{m\omega_m^2 \hat x^2}{2} +\frac{m\omega_{\rm SN}^2 (\hat x - \langle \hat x\rangle)^2}{2}
\end{equation}
with measurement record given by
\begin{equation}
y  dt =\alpha \langle \hat x \rangle dt +\frac{dW}{\sqrt{2}}
\end{equation}
Here $\langle \cdot \rangle$ signifies that we take the expectation value using $\rho$, with
\begin{equation}
\langle x \rangle =\mathrm{tr} (\hat x \hat\rho)
\end{equation}
Here $dW$ is a Wiener increment, and the Stochastic differential equations follow the Ito rule, 
\begin{equation}
dW^2 = dt\,.
\end{equation}
The formulation here, except the last term of Eq.~\eqref{SN_H}, follows the standard ``quantum trajectory'' treatment of continuous position measurement, which can be found in the textbook \cite{WM99} and the review article \cite{yanbei13b}.

For Gaussian states, which are completely characterized by expectation values $\langle \hat x\rangle$, $\langle \hat p\rangle$ and the covariance matrix components $V_{xx}$, $V_{xp}$ and $V_{pp}$, there is a self-contained set of equations for these first- and second moments \cite{yanbei23}.  At steady state, for the second moments, we have
\begin{align}
V_{xx}& =\frac{\hbar}{\sqrt{2} M\Omega}\frac{1}{\sqrt{1+\sqrt{1+\Lambda^4}}} \\
V_{xp} =V_{px} &=\frac{\hbar}{2}\frac{\Lambda^2}{1+\sqrt{1+\Lambda^4}}\\
V_{pp} &=\frac{M\Omega}{\sqrt{2}}\frac{\sqrt{1+\Lambda^4}}{1+\sqrt{1+\Lambda^4}}
\end{align}
Here we have defined \begin{equation}
\Omega =\sqrt{\omega_m^2  +\omega_{\rm SN}^2}\,,\quad
\Lambda=\sqrt{\frac{\hbar\alpha^2}{M\Omega^2}}\,.
\end{equation}
Interestingly, these second moments are the same as those obtained in standard quantum mechanics --- if the oscillator's mechanical resonant frequency were up-shifted to $\Omega$.  This is consistent with the intuition that in semiclassical gravity, the ``self-interaction'' term increases the oscillation frequency of the oscillator's quantum uncertainty, as indicated in Fig.~1 of Ref.~\cite{yanbei13a}.

In the limit of very weak measurement, with $\Lambda \ll 1$, this is the ground state of a harmonic oscillator with eigenfrequency $\Omega$.
 This is a squeezed state with
\begin{equation}
V_{xx} =\frac{\omega_m}{\Omega}  V_{xx}^{\rm vac}\,,\quad
V_{xx} =\frac{\omega_m}{\Omega}  V_{xx}^{\rm vac}\,.
\end{equation}
This state has a squeeze factor of
\begin{equation}
e^{2r} = \frac{\omega_m}{\Omega}
\end{equation}

\section{Correlated Worldline Theory}
 \label{sec:CWL}

We now turn to the CWL theory. We begin with a brief description of its salient features, followed by a discussion of how one can treat extended massive bodies. In the next section we deal with the specific case of a massive oscillator.

\subsection{Key Results from CWL Theory}
 \label{sec:3.A-CWLbasic}

Introductory discussions of CWL theory have appeared in a few places, either as general overviews \cite{carney19}, or in the introductory sections of more specialized papers \cite{CWL3,stamp15}. In this section we very briefly recall those features of CWL theory relevant to this study.

\subsubsection{Propagator in CWL Theory}
 \label{sec:3.A.1-CWL-K}

CWL theory is formulated in terms of paths (for particles) or `histories' (for fields). Thus one starts from Feynman paths; recall that the conventional non-relativistic Feynman propagator $K_o(2,1)$ for a particle with action $S_o[Q]$, between states $|1 \rangle \equiv |Q_1 \rangle$ and $|2 \rangle \equiv |Q_2 \rangle$ at times $t_1$ and $t_2$, is
\begin{equation}
K_o(2,1) \;=\; \int_1^2 {\cal D} Q(t)\, e^{iS_{o}[Q]}
 \label{Kqm-SHO}
\end{equation}
in which one sums over all different possible paths for the system, with each path contributing independently to the sum.

In the CWL theory one uses the same action, but there is a key difference in the evaluation of path integrals; now one has interactions between all the paths, mediated by the spacetime metric field $g^{\mu\nu}(x)$ (which we will often write as $g$). The CWL propagator ${\cal K}(2,1)$ for a particle is then found by labelling each of the different paths $q_k$ in a multiplet of $N$ paths by the index $k = 1,2, \cdots N$. One then defines the propagator for the combined SHO + metric field system between 2 spacetime hypersurfaces $\Sigma_1$ and $\Sigma_2$, as:
  \begin{widetext}
\begin{equation}
 \label{K21-CWL}
{\cal K}(2,1) \; = \; \lim\limits_{N\rightarrow\infty}\; \left({\cal K}_N(2,1) \right)^{1/N}  \;\;\;  = \;\;\;\;  \lim\limits_{N\rightarrow\infty}\; \left[\mathcal{N}_{N}^{-1}  \int^{\mathfrak{h}_2}_{\mathfrak{h}_1}  \mathcal{D}g\,e^{iNS_{G} [g]}  \prod_{k=1}^{N}\int_{Q_{1}}^{Q_{2}}\mathcal{D} q_{k}\, e^{iS_{o}[q_{k},\,g]} \right]^{1/N}
\end{equation}
	\end{widetext}
in which  $\mathfrak{h}^{ab}_1, \mathfrak{h}^{ab}_2$, are the initial and final induced metrics, specified on the initial and final hypersurfaces $\Sigma_1$ and $\Sigma_2$ respectively. Henceforth we will simple take these hypersurfaces to be time slices at times $t_1$ and $t_2$. The factor $\mathcal{N}_{N}$ is a normalization factor which we will not need.

The action $S_o[q_k, g]$ in (\ref{K21-CWL}) is written for a point-particle. In the next sub-section we discuss how things work for an extended mass. More generally CWL theory can be used for a set of quantum matter fields, in which case the paths are replaced by `histories', ie., the functional integration over different configurations of the fields \cite{CWL3}. 

As discussed in some detail in ref. \cite{CWL3}, the expression (\ref{K21-CWL}) can be evaluated both in exact form and as a perturbation expansion in the Newtonian coupling $G_N$; both forms are useful. The exact result, up to an overall renormalization, is
	\begin{equation}
\mathcal{K}(2,1)=e^{i(S_{G}[\bar{g}_{21}] +\psi_{0}(2,1|\bar{g}_{21}))},
   \label{K-eik}
\end{equation}
in which $\bar{g}_{21}$ is that metric which satisfies the conditional stationary phase requirement
\begin{equation}
	\label{eq:semiclassprop}
		\frac{\delta}{\delta g}\bigg(S_{G}[g] +\psi_{0}(2,1|g)\bigg)\bigg|_{g=\bar{g}_{21}}=0.
\end{equation}
ie., it is the solution to this differential equation with the metric $\bar{g}(x)$ subject to the boundary condition that the induced metrics on $\Sigma_1$ and $\Sigma_2$ are $\mathfrak{h}^{ab}_1$ and $\mathfrak{h}^{ab}_2$ as before.  The phase $ \psi_{0}(Q_{2},Q_{1}|g_{0})$ appearing in this equation is just the logarithm of a conventional propagator, given for a system with coordinate $Q$ by
\begin{equation}
K_0(Q_2, Q_1 |g_{0}) \;=\; e^{i\psi_{0}(Q_{2},Q_{1}|g_{0})}
	\label{K-phi-12'}
\end{equation}
where $g_0$ is the background spacetime. 

One then finds that
	\begin{align}
		\frac{\delta}{\delta g^{\mu\nu}(x)}\psi_{0}(2,1|g)\; &=\;  -\tfrac{1}{2}\frac{\langle Q_2|T_{\mu\nu}[x|g]|Q_1\rangle}{\langle Q_2|Q_1\rangle} \nonumber\\
& \equiv \;  -\tfrac{1}{2} \mathbb{\chi}^{\mathbb{T}}_{\mu\nu}(2,1|x,g)
	\label{T21-derive}
\end{align}
which defines the quantity $\mathbb{\chi}^{\mathbb{T}}_{\mu\nu}(2,1|x,g)$, which is basically a complex-valued matrix element. We think of it as a ``conditional stress-energy", ie., the stress energy $T_{\mu\nu}(x)$, subject to the condition that $Q(t)$ propagates between $Q_1$ on $\Sigma_1$ (ie., at time $t_1$) and  and $Q_2$ on $\Sigma_2$ at time $t_2$, on a background metric $g$. Another way to think of $\mathbb{\chi}^{\mathbb{T}}_{\mu\nu}(2,1|x,g)$ is as an expectation value of $T_{\mu\nu}(x)$, but one which is conditional on the pre- and post-selection of states $|Q_1 \rangle$ and $|Q_2 \rangle$ respectively (what is often referred to as a ``weak measurement" \cite{ABL64,wkMmt}).

The first functional derivative $\delta S_G/\delta g$ is of course that which normally defines the Einstein tensor $G_{\mu\nu}$, as it appears in the equation of motion $G_{\mu\nu}(x)=\ 8\pi G_N T_{\mu\nu}(x)$. However here we have instead that
\begin{equation}
		G_{\mu\nu}(\bar{g}_{21}(x))\;\;=\;\; 8\pi G_N \, \mathbb{\chi}^{\mathbb{T}}_{\mu\nu}(2,1|x,\bar{g}_{21})
	\label{KG21-KT21}
\end{equation}
in which the left- and right-hand sides are complex valued.

Turning now to the perturbative expansion of $\mathcal{K}(2,1)$, we show in Fig. \ref{fig:CWL-lowOrder} the first few terms; for a detailed discussion see ref. \cite{CWL3}. We emphasize two points here:


\begin{figure}[ht]
\centering
\includegraphics[scale=0.35]{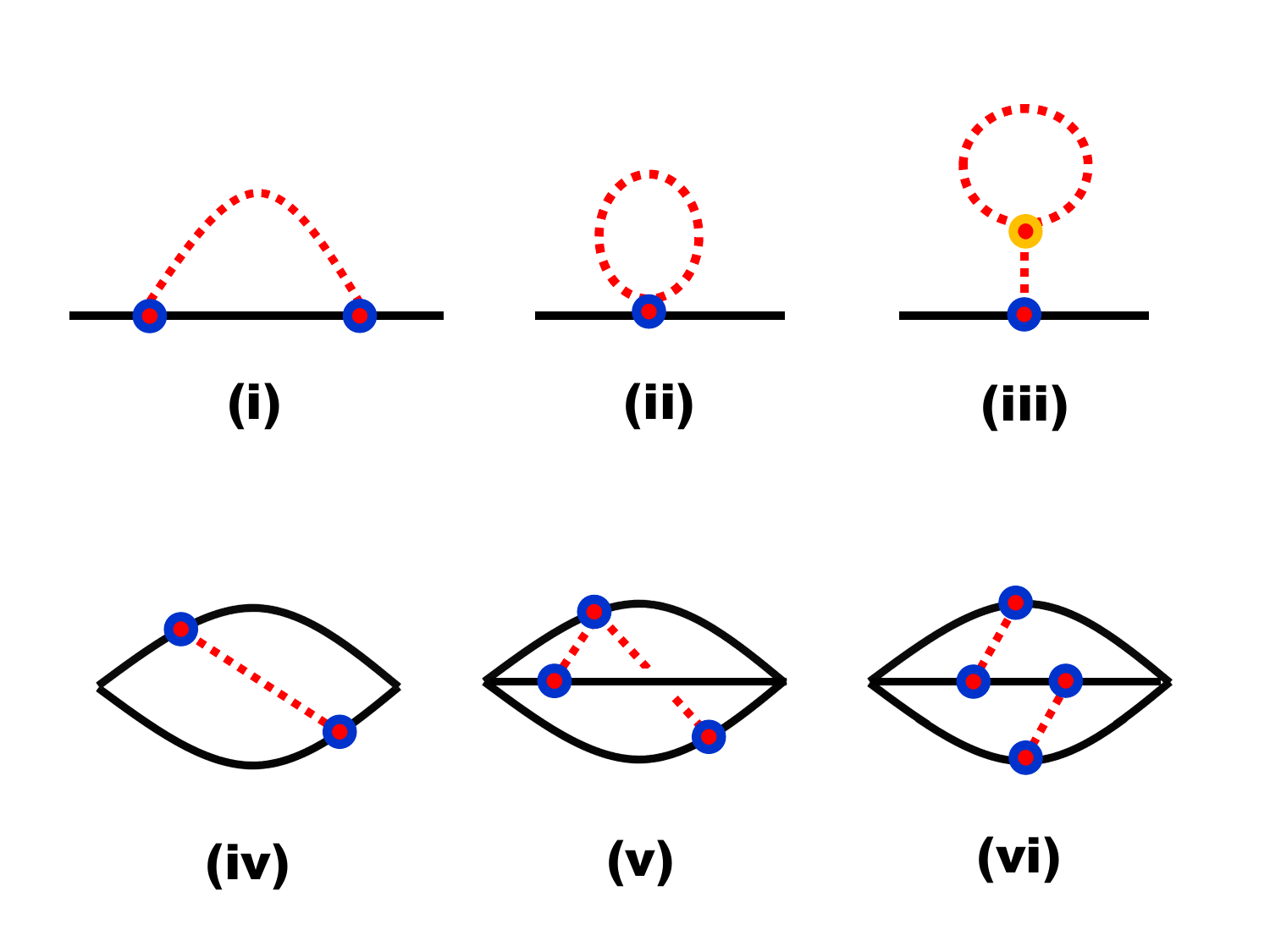}
\vspace{-7mm}
\caption{ Low-order contribution to the propagator ${\cal K}(2,1)$ in CWL theory; the thick dark lines represent the oscillator coordinate $Q$, an the hatched lines represent gravitons. In CWL theory, graph (i) is a `rainbow' graph, (ii) is a tadpole graph, and (iii) incorporates a 3-graviton vertex. Graph (iv) describes graviton exchange between a pair of paths for the oscillator, and graphs (v) and (vi) exchange gravitons between triplets of paths. Graphs (i)-(iii) give zero contribution in CWL theory, whereas graphs (iv)-(vi) all give finite contributions. }
 \label{fig:CWL-lowOrder}
\end{figure}


(i) the key physical point is that there are now attractive gravitational interactions between all paths of a given particle or matter field. In general one expects these interactions will cause all the paths to ``bunch" together, provided one can get rid of the kinetic energy stored up in relative motions between them - we will often in this paper refer to this as the `internal' CWL energy. However, this bunching interaction is negligible unless the mass $M$ of the object is sufficiently large \cite{CWL3,stamp15}. 

There are several ways that path bunching can occur. If, for example, the matter system is coupled to a cold environment, then the energy in the matter paths will be dissipated into this environment, and the loss of kinetic energy in the matter paths then allows them to bunch. Note however that if the environment is hot, the opposite can happen. 

(ii) Many of diagrams that one sees in conventional quantum gravity are actually absent in CWL theory. For example, the `rainbow graph' in Fig. \ref{fig:CWL-lowOrder}, which in conventional theory gives the lowest-order radiative correction to the dynamics of the mass $M$, is absent. As a general rule, there is no contribution in CWL theory from any diagram in which a loop integral contains a graviton line. 

However, one finds that if path bunching has occurred, all the conventional diagrams are restored \cite{CWL3}, since the separate paths now collapse into one single `semiclassical' path. In this way the interactions between separate paths are transformed into self-interactions for the classical path.

\subsubsection{Extended Mass in CWL Theory}
 \label{sec:3.A.2-CWLextendedM}

All of the above results are formal. We now discuss what happens when one applies them to the motion of a non-relativistic extended solid mass. It is clear from the lowest-order calculations done in CWL theory \cite{CWL3,stamp15}, up to $\sim O(G_N)$, that for masses $M \sim 10^{-17}$ kg or greater, one has to do all CWL calculations for an extended mass. This mass is far less than the Planck mass $M_p \sim 2 \times 10^{-8}$ kg. An object of mass $\sim 10^{-17}$ kg will have linear dimension $\sim 1.5 \times 10^{-7}$m, and (for SiO$_2$) contain $\sim 3 \times 10^8$ ions.

We also note that this mass may have an internal structure ranging from entirely crystalline order to one that is completely amorphous. The question at issue is then - what is the effective CWL interaction between different paths for the centre of mass motion of this mass? 

This question has been examined in detail elsewhere \cite{jordanPhD,CWL4}; here we just summarize the results which will be needed in the rest of this paper. 

We first note that almost all of the mass of the object is contained in the nuclei. To calculate the CWL interaction between 2 paths of the entire mass, one begins by describing the extended mass by the standard non-relativistic action
\begin{equation}
S_o[ \{ {\bf q}_j \}] \;=\; \int d\tau \left[ \sum_{j=1}^{\tilde{N}} {m_j \over 2} \dot{{\bf q}}_j^2 - \sum_{i<j}^{\tilde{N}} V({\bf q}_i - {\bf q}_j) \right]
 \label{effS-manyP}
\end{equation}
where the 3-vectors ${\bf q}_j(t)$ label the positions of each of the $\tilde{N}$ ions as a function of time $t$, and the $\{ m_j \}$ are the masses of the nuclei at sites $\{ j \}$ in the system. 

We now define the relative coordinates ${\bf r}_j(t)$ by ${\bf q}_j = {\bf R_o} + {\bf r}_j$, where
\begin{equation}
{\bf R_o}(t) = {1 \over {\tilde{N}}} \sum_{j=1}^{\tilde{N}} {\bf q}_j(t)
 \label{com}
\end{equation}
is the centre of mass of the body (so that $\sum_{j=1}^{\tilde{N}} {\bf r}_j = 0$). 

In ordinary QM the dynamics of this system is then obtained, in
path integral theory, by separating out the centre of mass and
rotational dynamics of the entire extended mass, from the internal
phonon modes \cite{jordanPhD,CWL4}. Let us here ignore rotational
modes modes here, and assume for the moment that the object is at low temperature, with $T\lesssim 50$K (we restore the effect of  thermal phonons later). One then
finds, as expected, that for a given centre of mass trajectory ${\bf
R}_o (t)$, which then determines the equilibrium paths ${\bf
\bar{r}}_j (t)$ for each nucleus, the paths of each individual
nucleus show quantum fluctuations in the quantum fluctuation coordinate ${\bf u}_j (t)$ around the
equilibrium paths (so that ${\bf r} = {\bf \bar{r}} + {\bf u}$).

At low $T$, these fluctuations are nothing but the zero-point phonons of the system. For a crystalline system these phonons have dispersion
\begin{equation}
\omega^2_{{\bf k} \mu} \;=\; \sum_{j} {1 \over (m_im_j)^{1/2}} V_{ij}\, e^{i {\bf k} \cdot {\bf \bar{r}}_{ij}}
 \label{omega-k}
\end{equation}
where $V_{ij} \equiv V({\bf \bar{r}}_i - {\bf \bar{r}}_j)$, and ${\bf \bar{r}}_{ij} \equiv {\bf \bar{r}}_i - {\bf \bar{r}}_i$; and the phonon correlator is
\begin{eqnarray}
\langle u_i^{\alpha}(t_1) u_j^{\beta} (t_2) \rangle \;&=&\; {1 \over {\tilde{N}}} \sum_{{\bf k} \mu} {1 \over (m_im_j)^{1/2}} \nonumber \\
&& \;\; \times {\hat{e}^{\alpha}_{{\bf k} \mu} \hat{e}^{\beta}_{{\bf k} \mu} \over 2 \omega_{{\bf k} \mu}} \, e^{i [{\bf k} \cdot {\bf \bar{r}}_{ij} - \omega_{{\bf k} \mu}(t_1 - t_2)]}
\label{phononC}
\end{eqnarray}

At low $T$ the zero-point amplitude $\xi_o$ of these fluctuations can be estimated in various ways - thus, in a simple Debye model one has $\xi_o \sim \tfrac{3}{2} \hbar (m \theta_D)^{-1/2}$, and more generally, one has $\xi_o \sim a_o E_{\Phi}/E_C$, where $a_o$ is the typical nearest neighbour distance between nuclei, $E_{\Phi}$ the characteristic elastic energy and $E_C$ the characteristic Coulomb energy in the solid. Typically one has $\xi_o \sim 1-3 \times 10^{-12}$m, ie., $\xi_o \sim 10^3$ nuclear radii.

Now in a CWL calculation to lowest-order in $G_N$, in which pairs of paths for the extended mass interact via non-relativistic Newtonian interactions, one must then evaluate a sum over all pairs of interactions of form
\begin{widetext}
\begin{equation}
V_{eff}[{\bf R}_o, {\bf R'}_o; \{ {\bf u}_{i}(t),  {\bf u'}_{j}(t)
\}] \;\;\;=\;\;\; -G_N \, \sum_{i,j=1}^{\tilde{N}} \; { m_{i}m_{j} \over |
({\bf R}_o(t)+{\bf \bar{r}}_{i}+{\bf u}_{i}(t)) \;-\; ({\bf R'}_o(t)+
{\bf \bar{r}}_{j}+ {\bf u'}_{j}(t))|}
 \label{extM-2path}
\end{equation}
\end{widetext}
and, once one has then path integrated over all pairs of paths $\{ {\bf u}_{i}(t)),  {\bf u'}_{j}(t)) \}$ for the phonon fluctuations, we get an effective interaction $V_{eff}({\bf R}_o, {\bf R}_o^{\prime})$ between pairs of paths for the centre of mass coordinate of our extended body.

The general results are quite complicated \cite{jordanPhD,CWL4}. Let us consider the case of an amorphous system, which is gives an approximate description of many of the mirrors used in optomechanical systems (for the discussion of different crystalline system see refs. \cite{jordanPhD,CWL4}). For simplicity we will assume $m_j = m$ for all the ions in the system.


\begin{figure}[ht]
\centering
\includegraphics[scale=0.35]{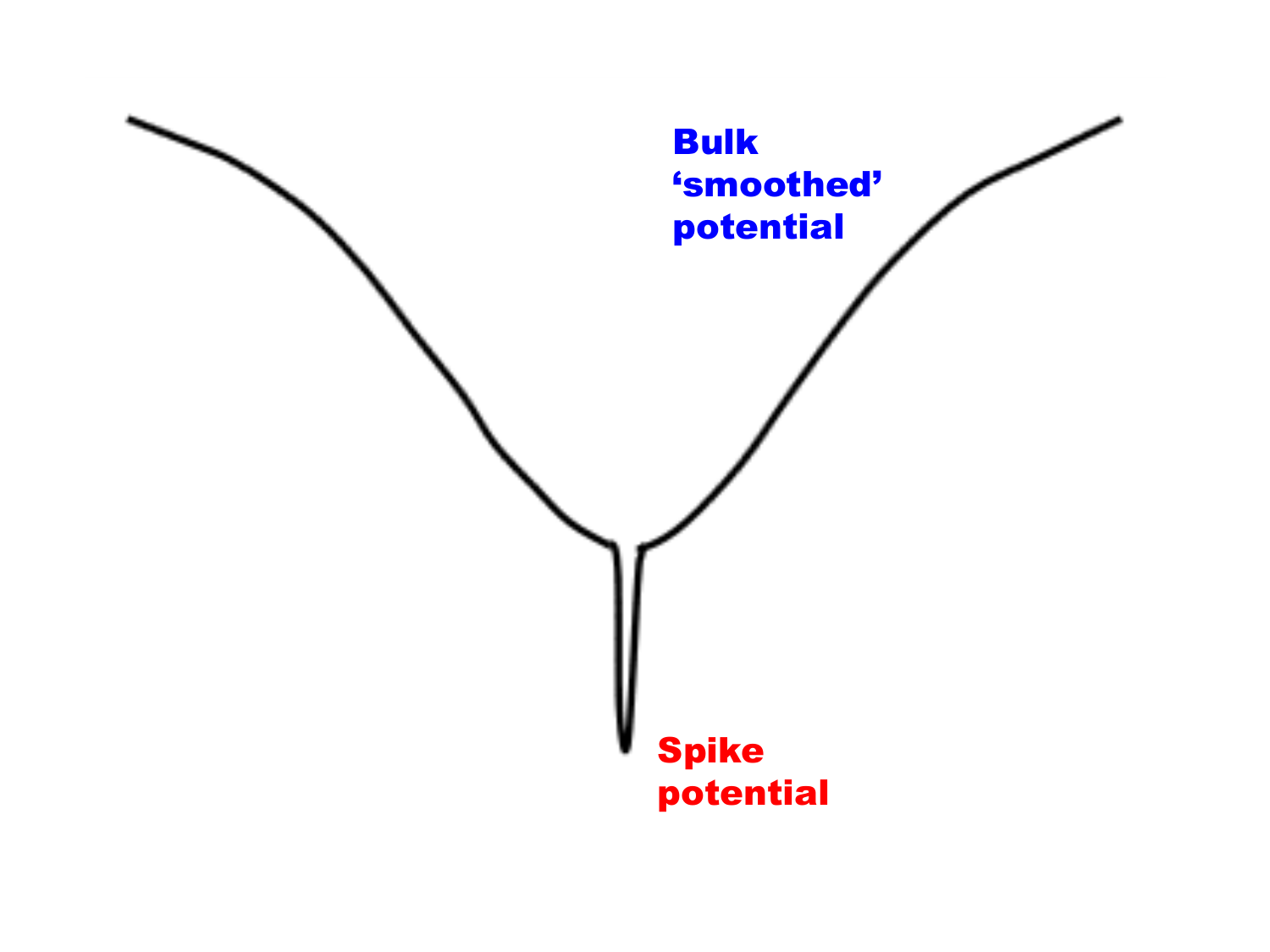}
\vspace{-7mm}
\caption{ The effective potential acting between the centre of mass coordinates of a pair of paths, for an extended massive body. The bulk smoothed potential is $V_{slow}(R)$ from eqn. (\ref{smoothfit}), and the `spike' potential is given in eqn. (\ref{spikefit}). The width and depth of the spike contribution are exaggerated.      }
 \label{fig:spike}
\end{figure}


We then get the result shown in Fig. \ref{fig:spike} for the effective CWL interaction, as a function of the distance $R = |{\bf R}_o - {\bf R'}_o|$. There is a slowly-varying part, fairly accurately described by
\begin{equation}
 \label{smoothfit}
V_{slow}(R) \;=\; -\frac{GM^{2}}{R}\, \textrm{Erf}\left(\frac{\sqrt{\pi}}{2}\gamma\frac{R}{a_{0}L_o}\right).
\end{equation}
where, as before, $L_o$ is the spatial extent of the massive body (in this case, the width of the mirror along the direction it moves), where $a_o$ is the typical lattice spacing, and where
\begin{equation}
\gamma \;=\;  V_o^{-5/3}\int_{V_o} d^{3}r \int_{V_o} d^{3}r' \frac{1}{|{\bf r} - {\bf r'}|}
  \label{gamma}
\end{equation}
is an $\mathcal{O}(1)$ constant characterizing the shape of the body (here $V_o = \pi R_o^2 L_o$ is the volume of the system). We also have, in the centre of the slowly-varying potential, a `spike' potential, which itself is accurately described by
\begin{equation}
 \label{spikefit}
V_{spike}(R) \;=\; -\frac{GMm}{R}\, \textrm{Erf}\left(\frac{R}{\sqrt{2}\xi_o}\right)
\end{equation}
with a characteristic range $\xi_o$.


\begin{figure}[ht]
\centering
\includegraphics[scale=0.35]{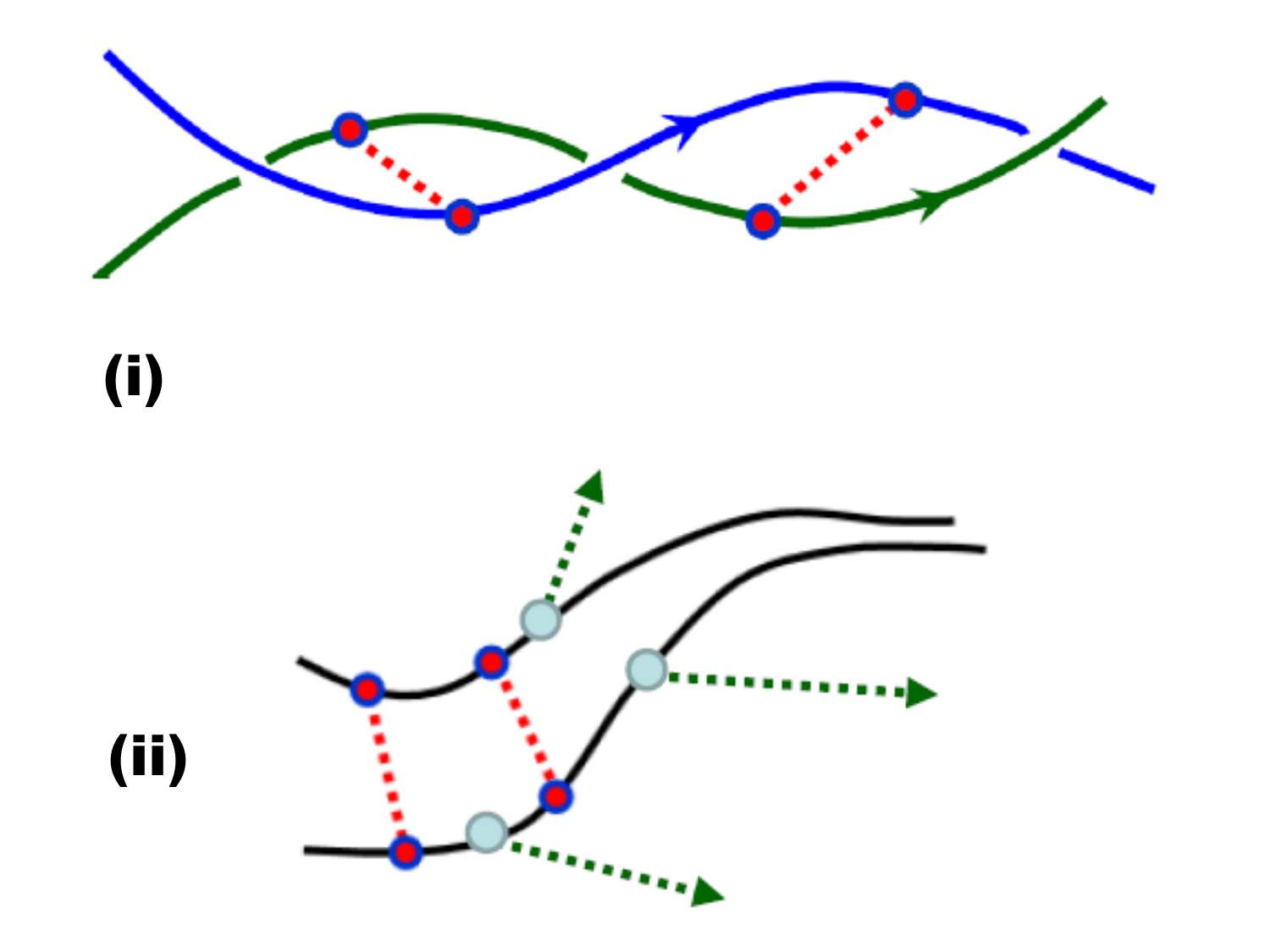}
\vspace{-7mm}
\caption{Trajectories of a pair of paths for a single object, under the influence of attractive gravitational CWL interactions (schematic).  In (i) these are shown in the absence of any path bunching, for a conservative system. In (ii) the paths are rapidly bunching as CWL gravitational energy is lost via emission of bath excitations. Solid lines are paths for the object, hatched lines are gravitons, and dotted lines are bath excitations.  }
 \label{fig:CWL-orbit}
\end{figure}


Because of this attractive interaction between pairs of paths, the paths will exhibit characteristic oscillations around each other (see Fig. \ref{fig:CWL-orbit}). The period of each of these oscillations depends on their amplitude:

\vspace{2mm}

(i) For the slowly-varying `bulk' part of the potential, one has
an oscillator frequency
\begin{equation}
\omega_B=\left(\frac{\pi}{6}\gamma^{3}G_N \rho_o \right)^{1/2}.
\end{equation}
where again  $\rho_o = M/ \pi R_o^2 L_o$ is the average mass density of the mirror. The time scale of these oscillations is long - thus, for amorphous SiO$_2$, one finds a period $\tau_B = 2\pi/\omega_B \sim 21.2$ mins. We note that the period is independent of the size of the mirror - this is because an increase in this size increases the strength of the coupling between the paths, but also increases the inertia associated with each path.
 \vspace{2mm}

 (ii) On the other hand the oscillation frequency inside the central `spike' potential is
\begin{equation}
 \label{spikefreq}
\omega_{sp}=\left(\sqrt{\frac{2}{\pi}}\frac{Gm}{3\xi_o^{3}}\right)^{\frac{1}{2}}.
\end{equation}
which is considerably higher; for amorphous SiO$_2$, we get a period $\tau_{sp} = 2\pi/\omega_{sp} \sim 16$ secs; again, this period is independent of the size of the mirror.

Finally, let us note that one can generalize these calculations to finite temperature $T$. For a general crystalline structure the results are rather complicated; however, for the isotropic amorphous system with a single ionic species just discussed, the modification is very simple. One finds that we must replace the factor $\xi_o$ in the above equations by a $T$-dependent $\xi_o(T)$, with 
\begin{equation}
\xi_o^2(T) \;=\; {\hbar^2 \over 2M} \int {dE \over E} g(E) [1 + 2 n(E,T)]
\end{equation}
in which $n(E,T)$ is the Bose distribution function, and $g(E)$ the phonon density of states. In a simple Debye model, with Debye energy $\theta_D$, this just gives
\begin{equation}
\xi_o^2 \;=\;  {9 \hbar^2 \over M} \left[ {1 \over 4\theta_D} + {1 \over \theta_D^3} \int_0^{\theta_D} dE {E \over e^{\beta \hbar E} - 1} \right]
\end{equation}
where $\beta = 1/kT$; this just reduces to the expression given previously as $T \rightarrow 0$. Note that while the `spike' potential is noticeably affected by a finite $T$, the slowly-varying bulk part is completely unaffected - this is because it is determined by the smoothed density of the system. 

\subsection{Factorizability of CWL States}
 \label{sec:CWL-factor}

A key motivation for discussing the \textit{pulsed} optomechanics experiment is the insensitivity of Gaussian oscillator states to CWL interactions. It is useful to give a simple explanation of why this is. Let's consider first a simple perturbative argument, to lowest order in the gravitational coupling, and suppose we have a simple harmonic oscillator (SHO) subject to some time-dependent force $F(t)$, with action
\begin{equation}
S_o[q|F] \;=\; \int dt \; [\tfrac{1}{2} m (\dot{q}^2 - \omega_m^2 q^2) - F(t)q(t) ]
 \label{So-SHO}
\end{equation}
with coordinate $z(t)$.

In CWL theory at lowest order in $G_N$ we simply deal with pairwise interactions between paths \cite{CWL3,stamp15}. We can then define `sum' and `difference' variables for the 2 paths $q_{1}(t)$ and $q_{2}(t)$ of the system, given by $Q = \tfrac{1}{2} (q_{1}+q_{2})$ and $z = \tfrac{1}{2} (q_{1}-q_{2})$.  Then then, up to $O(G_{N})$, the propagator for the oscillator is
\begin{eqnarray}
{\cal K}(2,1|F) \;&=&\; K^{-1}_0(2,1|F) \; \int_{1}^{2} {\cal D}Q \int_{0}^{0} {\cal D}z \, \times  \nonumber \\ &&e^{{i \over \hbar} (S_0[Q+z|F]+S_0[Q-z|F])}\; \exp^{i S_{CWL} [z]} \qquad
 \label{Ko-Qq}
\end{eqnarray}
where $S_{CWL}[z]$ is the CWL action representing the coupling between the pair of paths $q_{1},q_{2}$. Note that there is no requirement in the following argument for the CWL interaction to be harmonic, as we have so far assumed for an extended body. We could, for example, choose a Newtonian form between pairs of paths for a point particle, for which one has
\begin{equation}
S_{CWL} \;=\; {G m^2 \over 2} \int_{t_1}^{t_2}  {dt \over |z(t)|}  
 \label{S-CWL-N}
\end{equation}
in which the paths for $Q(t)$ have boundary conditions $Q(t_1) = Q_1$ and $Q(t_2) = Q_2$, while the paths for $z(t)$ have boundary conditions $z(t_1) = z(t_2) = 0$.

It is immediately clear that in the case where we have a quadratic form in $q$ and $\dot{q}$ for the action $S_o[q|F]$, as given in (\ref{So-SHO}), then the double path integral in (\ref{Ko-Qq}) factorizes \cite{stamp15}. The result is then
\begin{equation}
{\cal K}(2,1|F)  =  {1 \over K_0(2,1|F)} \int_{1}^{2} {\cal D}Q  \;
e^{{i \over \hbar} S_+[Q|F]}     \int_{0}^{0} {\cal D}z \; e^{{i \over \hbar} S_-[z]} \qquad
 \label{K-sep}
\end{equation}
where the two action terms are
\begin{align}
&S_+[Q|F] = 2 \int dt \; [\tfrac{1}{2} m (\dot{Q}^2 - \omega_m^2 Q^2) - F(t)Q(t) ] \qquad \nonumber \\
&S_-[z]= 2 \int dt \; \tfrac{1}{2} m (\dot{z}^2 - \omega_m^2 z^2) + {G m^2 \over 2 } \int  {dt \over |z(t)|}
 \label{S+-}
\end{align}
in which the time integrals are taken between $t_1$ and $t_2$.

eqn. (\ref{S+-}) yields a  propagator ${\cal K}(2,1|F)$, of factorized form, ie., we have
\begin{equation}
{\cal K}(2,1|F) \; \;=  K_0(2,1|F) K_C(0,0)
 \label{K-fact}
\end{equation}
where the first term $K_0(2,1|F)$ is the conventional QM propagator for a driven  oscillator, and where $K_C(0,0)$ describes the `path-bunching' dynamics of the relative coordinate $z(t)$ between the 2 paths, here in the Newtonian limit. 

This argument generalizes to the full situation with $N$ paths. The path integral for a CWL propagator contains the action
\begin{equation}
S_{eff} \;=\; \sum_k S_o[q_k] + \sum_{k,k'} S_{CWL} [q_k - q_{k'}]
 \label{Seff-CWL}
\end{equation}
and if the system is an oscillator we have, as before, that
\begin{align}
\sum_k S_o[q_k] =&\;  \frac{m}{2} \int dt \,\,\, N (\dot{Q}^2 - \omega_m^2 Q^2) \nonumber \\
+& \sum_k \int dt\, (\dot{z}_k^2 - \omega_m^2 z_k^2)
 \label{So-kz}
\end{align}
where $Q = N^{-1} \sum_k q_k$ and $z_k = q_k - Q$ are the new sum and difference coordinates (compare, eg., eqn. (\ref{S+-})). Thus again the path integral factorizes.

We can now conclude the proof of our claim from section 2.A. For such a factorized path-integral the `sum' coordinate decouples from the `difference' coordinates. If the initial state is Gaussian, it too will factorize into a Gaussian for the sum coordinate and Gaussians for the difference coordinates. Evolution under the factorized propagator preserves this form and the final state continues to be factorized. From here follows the main point, the CWL interactions occur only betweeen the difference coordinates but a linear measurement (ie. of the position of the particle) is only sensitive to the sum coordinate. This is a simple fact about how the different paths in CWL do not describe different particles, rather they are different paths (or \textit{histories}) of the same particle. For non-linear measurements or non-Gaussian states the factorization between sum and difference coordinates is spoiled and the CWL interactions play a non-trivial role. However, for linear measurements and Gaussian states the CWL effects simply decouple.

\section{Oscillators in CWL Theory}
 \label{sec:3.A.3-CWLstates}

In traditional QM one defines quantum states at specific times; in QFT one can define state functionals on specific hypersurfaces (which in flat spacetime are usually chosen to be time slices). In CWL theory, one can adapt the standard QFT treatment to give a definition of states on a spacelike hypersurface. 

Rather than give a general treatment, we simply work things out for the case of a harmonic oscillator. One can give explicit results for this case, and it is of course directly related to the optomechanical system we are dealing with.

\subsection{CWL Oscillator States}
 \label{osc-CWL}

The conventional action for the SHO is just
\begin{equation}
S_o \;=\; \tfrac{1}{2} \int dt M( \dot{Q}^2 - \omega_m Q^2)
 \label{SHO}
 \end{equation}
and in what follows we will define the notion of ``state" for the SHO, focusing on the ground state, and then we will show how to calculate this in CWL theory.

\subsubsection{Ground state of SHO}
 \label{sec:3.B.1-SHO-grdS}

{\bf (a) Conventional QM}: For conventional QM we define the evolution of a quantum state $|\psi(t)\rangle$ in the usual way, 
\begin{equation}
|\psi(t)\rangle \;=\; e^{-iHt}|\psi\rangle\; =\; \sum_{n}c_{n}e^{-iE_{n}t}|\phi_{n}\rangle.
\end{equation}
where the $\phi_n$ are the eigenstates of the Hamiltonian.  One can formally extract the ground state of this system by rotating to Euclidean time, ie., let  $t=-i\tau$,  so that \cite{coleman}
\begin{equation}
    \lim_{\tau\rightarrow\infty}|\psi(\tau)\rangle=\lim_{\tau\rightarrow\infty}\sum_{n}c_{n}e^{-\tau E_{n}}|\psi_{n}\rangle=c_{0}|\psi_{0}\rangle,
\end{equation}
where we've assumed $E_{0}=0$ and a unique ground state. For a any state such that $(c_{0}\neq 0$, we then have the ground state wavefunction
\begin{equation}
    \psi_{0}(Q)\propto \lim_{\tau\rightarrow\infty} \langle Q| \psi(\tau)\rangle.
\end{equation}

In a path integral treatment, this simply means that we have a
ground state wave-function given (up to normalization), for a system whose Lagrangian is a function of a coordinate ${\bf Q}$, by
\begin{equation}
    \psi_{0}({\bf Q}) \;=\; \int^{\bf Q}\mathcal{D}{\bf q} \,e^{-S_{E}[{\bf q}]}
    \label{psi0-PI}
\end{equation}
where the \textit{Euclidean action} is produced by the same rotation to imaginary time. Thus, eg., for a Lagrangian of form
\begin{equation}
L \;=\; \frac{1}{2}m^{jk}\frac{dq_{j}}{dt}\frac{dq_{k}}{dt}-V(\{q_{j}\})
\end{equation}
we get the Euclidean action
\begin{equation}
S_{E}[\{q_{j}\}]=\int^{0}_{-\infty}d\tau\,\left[\frac{1}{2}m^{jk}\frac{dq_{j}}{d\tau}\frac{dq_{k}}{d\tau}+V(\{q_{j}\})\right].
\end{equation}

Note that we leave the initial condition for each $q_{j}$ unspecified because it is clear that the final result is insensitive to this data. It is often convenient to choose $q_{j}(-\infty)=0$.

\vspace{4mm}

{\bf (b) CWL states}: Consider the general result given in eqn. (\ref{K21-CWL}) for the propagator ${\cal K}(2,1)$ in CWL theory, between states $1 \equiv Q_1$ at time $t_1$ and $2 \equiv Q_2$ at time $t_2$. Let us generalize this somewhat so that the system propagates between arbitrary states $|\alpha \rangle$ and $|\beta \rangle$, and also carry out the functional integration in (\ref{K21-CWL}) over the metric field $g^{\mu\nu}(x)$.

We are then left with
\begin{equation}
    {\cal K}(\beta, \alpha) \;=\; \lim_{N\rightarrow\infty}\left[\int^{\beta}_{\alpha}\mathcal{D}q_{1}...\int^{\beta}_{\alpha}\mathcal{D}q_{N}\,e^{iS[\{q_{k}\}]}\right]^{\frac{1}{N}}
     \label{Kba-CWL}
\end{equation}
where the effective (non-relativistic) CWL action for our SHO contains not only the oscillator potential in eqn. (\ref{SHO}), which acts equally on all of the CWL paths and which has frequency $\omega_m$, but also the CWL interactions between all of the paths. We will assume this interaction to be that derived in the last section, containing the slowly-varying term $V_{slow}(q_i-q_j)$ in eqn. (\ref{smoothfit}), along with the spike term $V_{spike}(q_i - q_j)$ in equation (\ref{spikefit}).

We will assume the system to be at very low $T$, and that the system has been prepared so that the paths have collapsed to the point where they are all confined to the inter-path spike potential (for what happens at finite temperature, see the end of this section).  In this case we will simply write the spike potential as an approximate harmonic potential well, with frequency $\omega_{\textrm{SN}}$ (where `SN' denotes `Schr\"odinger-Newton'). The assumption is that $\omega_{\textrm{SN}} \sim \omega_{sp}$, ie., that value given in eqn. (\ref{spikefreq}) for the spike oscillation frequency is a good approximation to the true value $\omega_{\textrm{SN}}$.

In this case we can write the effective CWL action in (\ref{Kba-CWL}) as
\begin{widetext}
\begin{equation}
    \label{eq:CWLevo}
    S \;\;=\;\; \frac{1}{2}\sum_{j=1}^{N}\int dt \left[M\left(\frac{dq_{j}}{dt}\right)^{2}-M\omega_{m}^{2}q_{j}^{2}-\frac{1}{2}\frac{M\omega^{2}_{SN}}{N}\sum_{k=1}^{N}(q_{j}-q_{k})^{2}\right].
\end{equation}
and the Euclidean form of this, which we shall use below, is then
\begin{equation}
    \label{eq:euclideanaction}
    S_{E} \;\;=\;\; \frac{1}{2}\sum_{j=1}^{N}\int_{-\infty}^{0} d\tau \left[M\left(\frac{dq_{j}}{d\tau}\right)^{2}+M\omega_m^{2}q_{j}^{2}+\frac{1}{2}\frac{M\omega^{2}_{SN}}{N}\sum_{k=1}^{N}(q_{j}-q_{k})^{2}\right]
\end{equation}
\end{widetext}

Returning now to our propagator ${\cal K}(\beta, \alpha)$, we notice that each of the initial states is the same (equal to $|\alpha \rangle$) and likewise for the final states (equal to $|\beta \rangle$); this is what we mean by a propagator.

\begin{figure}[ht]
\centering
\includegraphics[scale=0.35]{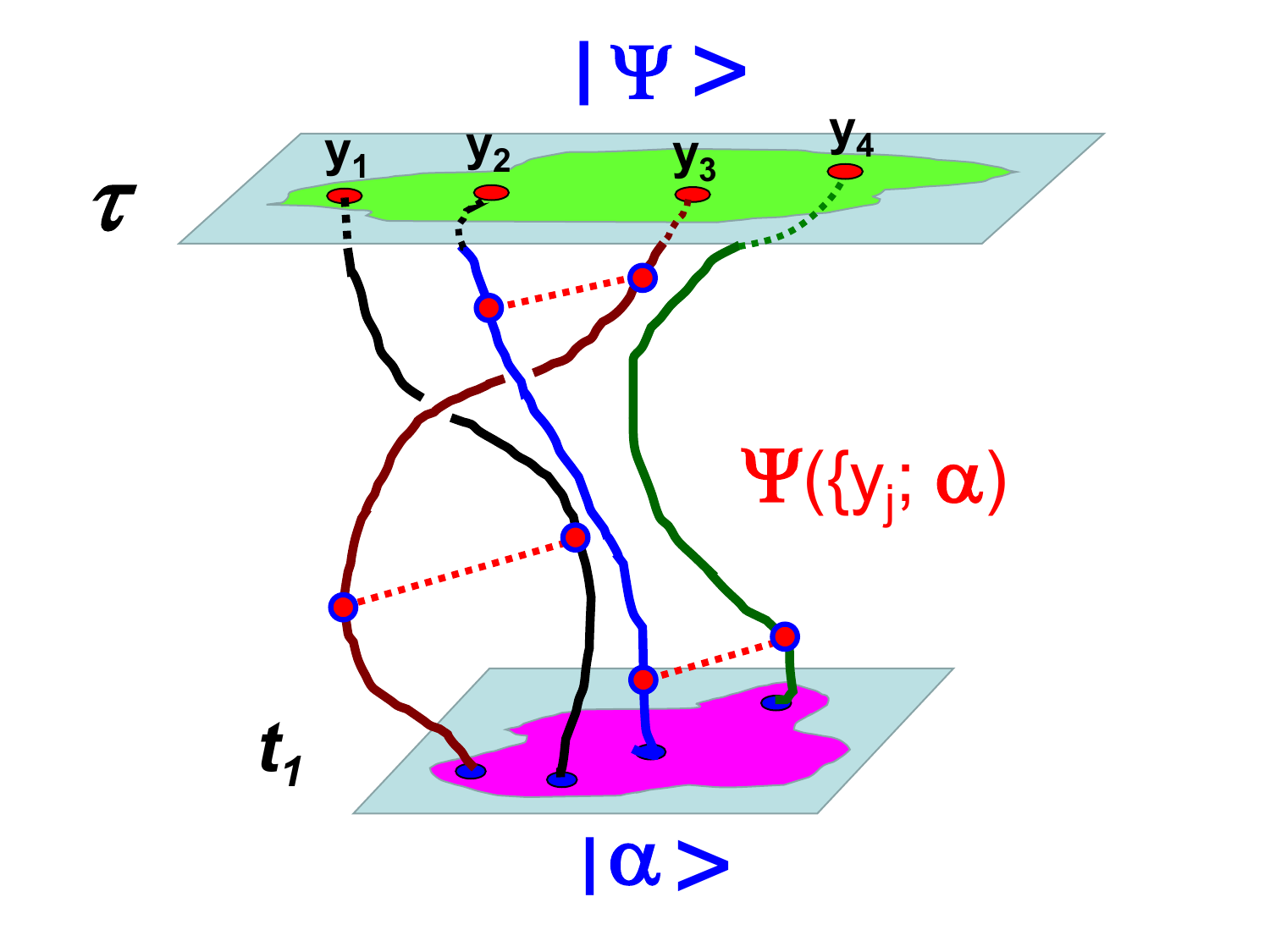}
\vspace{-7mm}
\caption{ The ``path multiplet" state functional $\Psi(\{y_{j}\}, \alpha)$. A physical system (here an oscillator with coordinate $q$) propagates from an initial state $|\alpha \rangle$ at time $t_1$ to a state $|\Psi \rangle$ at time $\tau$. In CWL theory one follows the fate of $N$ different paths for this object, which includes CWL interactions between the paths, thereby correlating them. The final position of the $j$-th path is $y_j$. Oscillator paths are shown as solid lines, gravitons interactions between these paths as hatched lines.  }
 \label{fig:psi-CWL}
\end{figure}

However suppose we wish to examine the state at some intermediate time (we assume here flat space). Then we can cut the propagator on the intermediate time slice ((for more details, see ref. \cite{CWL3}). We can then equivalently write the propagator as
\begin{widetext}
\begin{equation}
    \label{eq:CWLwithcorrelatedstate}
{\cal K}(\beta, \alpha) \;=\; \lim_{N\rightarrow\infty} \left[\int dy_{1}...dy_{N}\,\Psi(\{y_{j}\}, \alpha)  \int^{\beta}_{y_{1}}\mathcal{D}q_{1}...\int^{\beta}_{y_{N}}\mathcal{D}q_{N}\,e^{iS[\{q_{j}\}]}\right]^{\frac{1}{N}},
\end{equation}
\end{widetext}
where the intermediate ``path multiplet" state function (for $N$ paths) is defined by
\begin{equation}\label{eq:intermediatestate}
\Psi(\{y_{j}\}, \alpha) \;=\; \int^{y_{1}}_{\alpha}\mathcal{D}q_{1}...\int^{y_{N}}_{\alpha}\mathcal{D}q_{N}\,e^{iS[\{q_{j}\}]}.
\end{equation}
In what follows we will sometimes abbreviate the term `path multiplet' to `pathlet' and refer to individual paths as ``replicas''. We can represent the function in (\ref{eq:intermediatestate}) in diagrammatic form as shown in Fig. \ref{fig:psi-CWL}.


Note that this intermediate state function $\Psi$ is in general quite complicated - the effect of the CWL gravitational interactions is to correlate the paths, and so the functional $\Psi$ will not factorize into a product over each path. Note also that we have defined $\Psi(\{y_{j}\}, \alpha)$ starting from the {\bf initial} state $|\alpha \rangle$; but there is nothing stopping us from also writing a functional $\Psi(\{y_{j}\}, \beta)$ which starts from the final state $|\beta \rangle$. Our choice simply makes the traditional choice of retarded boundary conditions.

Finally, note that the intermediate state function is not an `observable' since all of the paths have different endpoints. Once computed, proper observables can be derived by integrating wave-functions against this functional. We will often find it useful to describe the evolution of these intermediate state functions via \textit{intermediate state propagators},
\begin{align}
    K(\{y_{j}\},\{x_{j}\})\;\;=\;\; \prod_{j}&\left(\int^{y_j}_{x_{j}}\mathcal{D}q_{j}\right)\,\, \exp^{iS[\{q_{j}\}]}\,,
\end{align}
where each replica $q_j$ may have its own start and end points independent of the other paths. The utility of this object is obvious, aside from the $N^\textrm{th}$ root in the definition of the CWL propagator we simply have a conventional unitary quantum mechanical system of N particles coupled via gravity. The properties of, and rules for manipulating,  these ``intermediate'' quantities are then identical to those in QM. Only at the end of the computation when we are to compute probabilities do we perform the novel-to-CWL $N^\textrm{th}$ root and encounter somewhat unfamiliar structures.

\subsection{Euclidean Time Evolution}
 \label{EuclidTE}

We now choose the intermediate state function to be a representation of the ground state of the system. That is, we write the Euclidean path integral
\begin{equation}
    \Psi_{0}(\{y_{j}\})=\int^{y_{1}}\mathcal{D}q_{1}...\int^{y_{N}}\mathcal{D}q_{N}\,e^{-S_{E}[\{q_{j}\}]}
     \label{Psi0-PI}
\end{equation}
in which we can now drop all reference to the initial state - this is precisely the manoeuvre we used in ordinary QM to define the ground state using a path integral (in eqn. (\ref{psi0-PI})). The Euclidean effective action in (\ref{Psi0-PI}) is just that given in eqn. (\ref{eq:euclideanaction}).

Let us now evaluate this function. Since the action is quadratic, the integral is Gaussian and can be performed exactly via saddle-point evaluation. The saddle-point equation, ie. classical equation of motion, is
\begin{equation}\label{eq:replicaeom}
    \frac{d^{2}{q}_{j}}{d\tau^{2}}=(\omega_m^{2}+\omega_{\textrm{SN}}^{2})q_{j}-\omega_{\textrm{SN}}^{2}Q,
\end{equation}
where we've defined the \textit{average} coordinate
\begin{equation}
    Q\equiv\frac{1}{N}\sum_{j}q_{j}.
\end{equation}
We can see in (\ref{eq:replicaeom}) that each replica behaves like a (Euclidean) simple harmonic oscillator, with effective frequency $\Omega\equiv\sqrt{\omega_m^{2}+\omega^{2}_{SN}}$, and driving force $F(t)=m\omega^{2}_{SN}Q(t)$ determined by the average motion of all replicas.

We can obtain an equation of motion for $Q$ by summing over the different pathlets in (\ref{eq:replicaeom}), and it is simply
\begin{equation}
    \frac{d^{2}Q}{d\tau^{2}}=\omega_m^{2}Q.
\end{equation}
The average pathlet coordinate for the oscillator system evolves without CWL corrections. We can immediately solve this equation for $Q$,
\begin{equation}
    \label{eq:Qsoln}
    Q(\tau)=Y\,e^{\omega_m\tau},
\end{equation}
where we've eliminated the solution which is divergent as $\tau\rightarrow-\infty$, and we've defined the average final coordinate $Y\equiv\frac{1}{N}\sum_{j}y_{j}$.

With this solution for $Q(\tau)$ in hand, the equation of motion (\ref{eq:replicaeom}) literally describes independent forced oscillators.

Given the simplicity we've found in the equation of motion, it is useful to go back to the action (\ref{eq:euclideanaction}) and rewrite it to demonstrate this factorization explicitly. To do so, note that the interaction can be written as
\begin{align}
    \frac{1}{N}\sum_{jk}(q_{j}-q_{k})^{2}&=2\sum_{j}q_{j}^{2}-2N\left(\frac{1}{N}\sum_{k}q_{k}\right)^{2} \nonumber \\
    &=2\sum_{j}q_{j}^{2}-2NQ^{2}.
\end{align}

We can insert this simple result directly into the action to write it as
\begin{widetext}
\begin{equation}
    S_{E}=\frac{-Nm\omega_{\textrm{SN}}^{2}}{2}\int^{0}_{-\infty} d\tau \, Q^{2}+\sum_{j}\int_{-\infty}^{0} d\tau \left[\frac{1}{2}m\left(\frac{dq_{j}}{d\tau}\right)^{2}+\frac{1}{2}m(\omega_m^{2}+\omega_{\textrm{SN}}^{2})q_{j}^{2}\right].
    \label{factor}
\end{equation}
Aside from the first term, the action is simply that of $N$ identical  oscillators.

The on-shell action, ie. that evaluated on the saddle-point solution, can be simplified further if we use the standard ``\textit{integrate-by-parts and apply the equation of motion}" trick to rewrite the kinetic term as
\begin{equation}
    \int_{-\infty}^{0}d\tau\left[\frac{1}{2}m\left(\frac{dq_{cl}}{d\tau}\right)^{2}\right]=\frac{m}{2}\left[q_{cl}\frac{dq_{cl}}{d\tau}\right]^{\tau=0}_{\tau=-\infty}-\int_{-\infty}^{0}d\tau\left[\frac{1}{2}m\Omega^{2}q_{cl}^{2}-\frac{1}{2}m\omega_{\textrm{SN}}^{2}Qq\right]
\end{equation}
The total on-shell action is then
\begin{equation}
    S_{E}=\frac{-Nm\omega_{\textrm{SN}}^{2}}{2}\int^{0}_{-\infty} d\tau \, Q^{2}+\sum_{j}\frac{m}{2}\left[q_{j\,cl}\frac{dq_{j\,cl}}{d\tau}\right]^{\tau=0}_{\tau=-\infty}+\frac{1}{2}m\omega_{\textrm{SN}}^{2}\sum_{j}\int_{-\infty}^{0}d\tau \; Q q_{j\,cl}.
\end{equation}
\end{widetext}

We now recognize that the final term, after evaluating the sum over $j$, exactly cancels the first term. The whole result for the on-shell action is simply
\begin{equation}\label{eq:onshellsimple}
S_{E}=\frac{m}{2}\sum_{j}\left[q_{j\,cl}\frac{dq_{j\,cl}}{d\tau}\right]^{\tau=0}_{\tau=-\infty}.
\end{equation}
It remains only to solve for $q_{j\,cl}$.

The general form of the solution to (\ref{eq:replicaeom}) is
\begin{equation}
    q_{j\,cl}(\tau)=a_{j}e^{\Omega\tau}+b_{j}e^{-\Omega\tau}-\omega^{2}_{SN}\int_{-\infty}^{0}d\tau'\,G(\tau,\tau ')Q(\tau'),
\end{equation}
where the Green's function is
\begin{equation}
G(\tau,\tau')=-\int_{-\infty}^{\infty}\frac{d\lambda}{2\pi}\,\frac{e^{-i\lambda(\tau-\tau')}}{\lambda^{2}+\Omega^{2}}=-\frac{e^{-\Omega|\tau-\tau'|}}{2\Omega}.
\end{equation}
Regularity as $\tau\rightarrow-\infty$ requires $b_{j}=0$, and we fix $a_{j}$ by imposing $q_{j}(0)=y_{j}$.  Using the explicit solution for $Q(\tau)$ in (\ref{eq:Qsoln}) we can then write down the remarkably simple solution for each replica's trajectory,
\begin{equation}
    q_{j\,cl}(\tau)=(y_{j}-Y)e^{\Omega\tau}+Ye^{\omega_m\tau}.
\end{equation}
Substituting this solution into equation (\ref{eq:onshellsimple}) for the on-shell action, we arrive at our final expression for the CWL ground state ``wavefunction''
\begin{equation}\label{eq:wfunction}
    \Psi_{0}(\{y_{j}\})=\exp\left(-\frac{1}{4}y_{j}y_{k}A^{jk}\right),
\end{equation}
with
\begin{equation}\label{eq:euclideanresult}
A^{jk}=2m\Omega\,\delta^{jk}-2m(\Omega-\omega_m)\frac{1}{N}\boldone^{j}\boldone^{k},
\end{equation}
where $\boldone^{j}$ is a vector of ones. For future applications it is often useful to write it as
\begin{equation}
A^{jk}=2m\omega_m\,e^{2\zeta}\,\delta^{jk}+2m\omega_m \frac{\delta}{N}\,\boldone^{j}\boldone^{k},
\end{equation}
where we've identified the effective squeezing parameter
\begin{equation}
    \zeta=\tfrac{1}{2}\ln\left(\frac{\Omega}{\omega_m}\right)=\frac{\omega^{2}_{SN}}{4\omega_m^{2}}+\mathcal{O}\left(\left(\tfrac{\omega_{\textrm{SN}}}{\omega_m}\right)^{4}\right),
\end{equation}
and the dimensionless CWL correlation parameter
\begin{equation}
    \delta=\frac{\omega_m-\Omega}{\omega_m}=-\frac{\omega_{\textrm{SN}}^{2}}{2\omega_m^{2}}+\mathcal{O}\left(\left(\tfrac{\omega_{\textrm{SN}}}{\omega_m}\right)^{4}\right).
\end{equation}

 One key limitation is that  the oscillator centre of mass coordinate is well within the ``spike'' potential shown in Fig. \ref{fig:spike}. As such they rely on our assumption of very low temperatures, which, as we discuss in the next section, will typically not apply for real experiments.

\section{Optomechanics in CWL Theory}
 \label{sec:4-optoM-CWL}

We now turn to a detailed description of our optomechanical system in CWL theory. To do this it is convenient to introduce a holomorphic (coherent state) description of the photon variables, using a coherent state path integral generalized to CWL theory. This work is done in preparation for the next section, where we show how one describes a pulsed optomechanics experiment of the kind described in the introduction.

We will be introducing coherent states for actions taking the general form
\begin{eqnarray}
    S[\{q_j\}] &\;=\; & \sum_{j}\int dt\left(\frac{m}{2}\dot{q_{j}}^{2}-V(q_{j})\right) \nonumber \\
    &\;&\;\;\;\;-\; \frac{1}{2N}\sum_{j\neq k}\int dt\, V^{CWL}(q_{j}-q_{k}) \;\;\;
\end{eqnarray}
in which $V(q_{j})$ is just the quadratic mechanical mirror potential:
\begin{equation}
    V(q_{j}) \;=\; \frac{m\omega_m^{2}}{2}q_{j}^{2}
\end{equation}
and where the CWL potential $V^{CWL}$ is also a quadratic form, describing the bottom of the `spike' potential introduced earlier:
\begin{equation}
    V^{CWL}(q_{j}-q_{k}) \;=\; \frac{m\omega_{\textrm{SN}}}{2}(q_{j}-q_{k})^{2}.
\end{equation}
Thus we are typically interested in very low temperatures, when the centre of mass motion of the system is confined to this spike potential (ie., to length scales $\sim 10^{-12}$m or less).

\subsection{Coherent State Representation}
 \label{sec:4.A-coherentS}

To evaluate the intermediate state propagator for the oscillator, and later for an optomechanical system,  it is convenient to utilize a coherent state representation.

To develop this, we first note that the CWL interaction has a useful simplifying feature; we can write
\begin{align}
    \frac{1}{2N}\sum_{j\neq k} \frac{m\omega_{\textrm{SN}}}{2}(q_{j}-q_{k})^{2}\;=\; \frac{m\omega_{\textrm{SN}}}{2} q_{j}q_{k}\left(\delta^{jk}-\frac{\boldone^{j}\boldone^{k}}{N}\right),
\end{align}
Using this notation, the CWL action reads
\begin{equation}
    S[\{q_j\}] \;=\;\int dt\left(\frac{m}{2}\dot{q}_{j}\dot{q}^{j}-\frac{m\Omega^{2}}{2}q_{j}q_{k}V^{jk}\right),
\end{equation}
where we've defined a matrix
\begin{equation}
    V^{jk} \;=\; \delta^{jk}-\frac{\omega^{2}_{\rm SN}}{\Omega^{2}}\frac{\boldone^{j}\boldone^{k}}{N},
\end{equation}
and we've defined the shifted frequency $\Omega^{2}=\omega_m^{2}+\omega_{\textrm{SN}}^{2}$.

In what follows we will be working only to leading order in the dimensionless CWL coupling strength $\epsilon^{2}=\omega^2_{\rm SN}/2\Omega^{2}$.

We will also make frequent use of the matrix
\begin{equation}
    P^{jk}\;=\; \delta^{jk}-\epsilon^{2}\frac{\boldone^{j}\boldone^{k}}{N},
\end{equation}
which to leading order is the square root of $V^{jk}$.

To move towards a a coherent state representation, we first integrate in the momentum variable.  That is, we multiply by
\begin{equation}
    1 \;=\; \prod_{j}\int\mathcal{D}p_{j}e^{-\frac{i}{2m}\int dt\,(p_{j}-m\dot{q}_{j})^{2}},
\end{equation}
which puts the intermediate state propagator into first-order form
\begin{widetext}
\begin{align}
    K(\{y_{j}\},\{x_{j}\})\;\;=\;\; \prod_{j}&\left(\int\mathcal{D}p_{j}\int^{y_j}_{x_{j}}\mathcal{D}q_{j}\right)\,\, \exp\left[i\int dt\left(p_{j}\dot{q}^{j}-\frac{p_{j}p^{j}}{2m}-\frac{m\Omega^{2}}{2}q_{j}q_{k}V^{jk}\right) \right]
\end{align}
\end{widetext}

In the present form we have fixed boundary data for the positions an unfixed boundary data for the momenta. We can however change basis quite easily by fixing eg, the momenta and not the positions, the initial positions and final momenta, or linear combinations of positions and momenta, provided we supplement the action with appropriate boundary terms so that the variational problem is well posed.

The coherent state representation follows from defining the new variables
\begin{align}
    &b_{j}=\sqrt{\frac{m\Omega}{2}}q_{j}+\frac{i}{\sqrt{2m\Omega}}P^{-1}_{jk}p^{k} \nonumber \\
    &\bar{b}_{j}=\sqrt{\frac{m\Omega}{2}}q_{j}-\frac{i}{\sqrt{2m\Omega}}P^{-1}_{jk}p^{k}
\end{align}
in terms of which the CWL action is simply
\begin{align}
    S[\{b_{j},\bar{b}_{j}\}]\;=&\; -i\bar{b}^{j}(t_{f})P_{jk}b^{k}(t_{f}) \nonumber \\ & + \int\,dt\left(i\bar{b}^{j}P_{jk}\dot{b}^{k}-\,\Omega\,\bar{b}^{j}P^{2}_{jk}b^{k}\right) \;\;
\end{align}

Since the equations of motion which follow from this action are first-order, we must fix mixed boundary data in the propagator in order for the variational problem to be well posed. We then get
\begin{widetext}
\begin{align}
    K(\{\bar{\beta}^{(f)}_{j}\},\{\beta^{(i)}_{j}\})\;\;=\;\; \prod_{j}\left(\int^{\bar{\beta}^{(f)}_j}\mathcal{D}\bar{b}_{j}\int_{\beta^{(i)}_{j}}\mathcal{D}b_{j}\right) \exp\left[\bar{\beta}^{(f)}{}^{j}P_{jk}b^{k}(t_{f})+i\int dt\left(i\bar{b}^{j}P_{jk}\dot{b}^{k}-\Omega\,\bar{b}^{j}P^{2}_{jk}b^{k}\right) \right]
\end{align}
\end{widetext}

The coherent state path-integral is almost trivial to evaluate, because the anti-holomorphic variables $\bar{b}^{j}$ appear only as Lagrange multipliers. We can thus immediately integrate them out, forcing the holomorphic variables $b^{j}$ to satisfy an equation of motion. The result is
\begin{align}
    K(\{\bar{\beta}^{(f)}_{j}\},\{\beta^{(i)}_{j}\}) \;\;=\;\; \exp\left[\bar{\beta}^{(f)}{}^{j}P_{jk}b_{cl}^{k}(t_{f},\{\beta^{(i)}_{j}\})\right]
\end{align}
where the $b_{cl}^{k}(t,\{\beta^{(i)}_{j}\})$ satisfy the first order equations of motion
\begin{equation}
    i\dot{b}^{k}-\Omega P_{jk}b^{j} \;=\; 0
\end{equation}
with the boundary conditions $b_{j}(0)=\beta^{(i)}_{j}$.

As an aside, we should note that these variables make it trivially quick to find the ground state of the CWL oscillator. The effective Hamiltonian can be read off as
\begin{equation}
    H\;=\;\Omega\,\bar{b}^{j}P^{2}_{jk}b^{k},
\end{equation}
and since the matrix $P^{2}$ is positive definite the ground state is the state for which all $b_{j}=0$. Translating this to an operator statement it reads
\begin{equation}
    \left(\frac{d}{dx^{j}}+m\Omega P_{jk}x^{k}\right)\Psi_{0}(\{x_{j}\}) \;=\; 0,
\end{equation}
which is satisfied by the wavefunction
\begin{equation}
    \Psi_{0}(\{x_{j}\}) \;=\; \exp\left[-\frac{m\Omega}{2}x^{j}P_{jk}x^{k}\right].
\end{equation}
This result agrees precisely with the wavefunction derived above (eqn. (\ref{eq:wfunction})) using a different method.

\subsection{Effect of Finite Temperature}
 \label{sec:4-finiteT}

All of the above calculations are essentially done in the limit where temperature $T \rightarrow 0$; physically, the assumption is the thermal energy $kT \ll |V_{spike}|$, the depth of the spike potential, ie., that the system is confined to the spike potential in eqn. (\ref{spikefit}). 

To see how realistic this is, let us consider an example. Suppose we have a mirror-shaped mass $M =$ 40 kg, made from SiO$_2$, with thickness $L_o =$ 16 cm, and diameter $d_o =$ 35 cm (this is the mass and shape of the advanced LIGO mirror). The we find the following results:

(i) The depth of the smoothed potential well is $|V_{smooth}| \sim GM^2/L_o \sim 5.17 \times 10^{12}$ eV $\equiv 6 \times 10^{16}$K, and the bulk small oscillation frequency is $\omega_B = 8.5 \times 10^{-4}$ Hz, so that the period of oscillation is $\tau_B = 2\pi/\omega_B \sim 123$ mins. 

(ii) The depth of the spike potential well is $|V_{spike}| \sim GM m_o/\xi_o$, where we will pick a mass $m_o \sim 30$ amu, and zero point length $\xi_o \sim 4 \times 10^{-13}$m. One then finds that $|V_{spike}| \sim 20.8 \times 10^{-4}$ eV $\sim$ 24K (ie., corresponding to a frequency $\sim$ 500 GHz).  The small oscillation frequency is then $\omega_{spike} \equiv \omega_{\textrm{SN}} = 0.37$ Hz, so that the oscillation period is $\tau_{SN} \sim 8.5$ secs. 

From these numbers we see that unless the system is at temperatures $\ll$ 24K, ie., cooled to liquid $^4$He temperatures or below, it will not be described by the zero-$T$ calculations just performed. At room temperature its motion will in fact hardly depend on the spike potential, and it will be in a high-$T$ state, oscillating rather slowly in the smooth bulk potential. 

From a purely theoretical point of view one can adopt two different approaches to the problem of describing the finite-$T$ motion. 

The first is to simply assume that not only is the entire system in thermal equilibrium, but the that the internal `kinetic' CWL degrees of freedom are also in thermal equilibrium with each other, at the same temperature. In this case one can redo the zero-$T$ calculations just done in a finite-$T$ generalization of the formalism. It is intuitively obvious that when $kT \gg |V_{spike}|$, the CWL oscillation frequency will drop from $\omega_{\textrm{SN}}$ down to $\omega_B$, and the magnitude of the corrections to conventional QM will then be $\sim (\omega_{B}/\omega_m)^2$, ie., far smaller again than even the very small corrections we have just derived for the $T=0$ case. 

The second way one can approach this problem is by looking at the `quantum relaxational dynamics' of the system, and in particular at the way in which the CWL internal kinetic degrees of freedom relax to the `central' QM dynamics. In fact this relaxation may be very slow, because they may be very weakly coupled to each other. Thus, for our mechanical mirror system, in many cases the internal CWL degrees of freedom $\{ {\bf q}_k \}$, associated with the $k$-th path, may be very far from equilibrium with the centre of mass coordinate ${\bf R}_o(t)$, and at a quite different temperature.  

One obvious way that this mutual equilibrium can be established is via the surrounding environment. To see how this happens, let's assume the mirror is coupled to an oscillator bath at a temperature $T$, with the oscillator Hamiltonian given in eqn. (\ref{2.3}). 

It is then possible to evaluate the dynamics of the oscillator centre of mass density matrix $\rho(Q,Q', t|\beta)$ as a function of time in CWL theory, by generalizing the work done in this paper. To do this is quite lengthy, so here we simply describe some of the conclusions:

(a) Without loss of generality we can keep only the coupling $F_q(P,Q)$ in the mirror-bath interaction \cite{ajl84}. We then find that in addition to this direct coupling between the mirror centre of mass coordinate $Q$ and the bath modes $\{ x_q \}$, there is also a coupling between all the internal CWL degrees of freedom (ie., the internal relative coordinates $q_k - q_{k'}$ between different paths $q_k, q_{k'}$ for the mirror) and the difference coordinates $x_q - x_q^{\prime}$ for the bath variables. 

(b) One can see how this works already by doing a calculation at leading order in perturbation theory, in which only the CWL interactions between pairs of paths are incorporated. To evaluate $\rho(Q,Q', t|\beta)$ one averages over the bath variables in the usual way. In CWL theory one then finds two influence functionals in $\rho(Q,Q', t|\beta)$, one for the `summed' (centre of mass) coordinate $(Q + Q')/2$, and the other for the difference coordinate $(Q - Q')/2$ which, at $\mathcal{O}(G_{N})$ represents the internal CWL mode. For the $F_q(P,Q)$ coupling just given, one finds this influence functional to be
\begin{eqnarray}
{\cal F}[u,u'] \;&=&\; \exp \int d\tau_1 \int d\tau_2 (u(\tau_2) - u(\tau_1)) \nonumber \\
&&  \times [{\cal D} (\tau_1 - \tau_2) u(\tau_2) - {\cal D}^* (\tau_1 - \tau_2) u'(\tau_2)] \;\;\;\;
 \label{F-uu'}
\end{eqnarray}
where the effective dynamic coupling, mediated by the bath, between the relative CWL coordinates, is
\begin{eqnarray}
{\cal D}(\tau) \;&=&\; \sum_q {1 \over 2 m_q \omega_q^2} \left( {\partial F_q(u) \over \partial u} \right)^2 \big{|}_{u = (Q - Q')/2} \nonumber \\
&& \qquad\qquad  \times \left[ e^{i \omega_q \tau} + 2 {\cos \omega_q \tau \over e^{\beta \omega_q} - 1} \right]
 \label{calD}
\end{eqnarray}
 
(c) As a consequence of this, the internal CWL modes will gradually equilibrate at the temperature of the bath. The timescale can be very long - as already noted previously \cite{CWL3}, if the characteristic frequency of relative oscillations between internal CWL modes is $\omega_{\textrm{SN}}$, and for a given $Q$-factor describing the oscillator decay when it is in contact with the bath, we find that a decay time $\tau_R \sim Q/\omega_{\textrm{SN}}$. Since $Q$ may be as high as $10^8$, this can give times of thousands of years. If however $Q \sim O(1)$, then the internal CWL degrees of freedom can relax to the bath temperature in times $\sim 10$ secs.

\medskip

\section{Pulsed Optomechanics Experiment}
 \label{sec:6-pulse-Opto}


We now return to the experimental protocol described in section 2, in which the rectangular pulse $g(t)$ in eqn. (\ref{pulseS}) is applied using the external laser. As a reminder, the idea we'd previously alluded to was to start with a mechanical oscillator in its ground state and the cavity mode in an excited state, apply a pulse to completely swap their states, wait a duration $T$, and then apply a pulse to completely swap their states back (see Fig. \ref{fig:cavity}). 

We will discuss first the CWL theory and then the SN theory - the case where ordinary QM applies will be obvious as the limit where $G_N \rightarrow 0$.

\subsection{Pulsed Optomechanics in CWL Theory}
 \label{sec:CWL-Opto}

To treat the pulsed system in CWL theory we first write down the expression for the CWL propagator for our general optomechanical system in coherent state representation, and then specialize to the case of the pulsed experiment described in the introduction. The most interesting thing to do is compare the state of the system before and after pulses, which we do by preparing the initial state of the mirror as a Fock state.

\subsubsection{CWL propagator}
 \label{sec:4.B.1-CWL-K-Opto}

We begin again with the general optomechanical Hamiltonian (\ref{eq:optomechhamiltonian}) given in section 2.
To write the propagator in CWL theory we write set up $N$-path multiplets for both the cavity mode and the mechanical oscillator, and couple them according to eqn. (\ref{eq:optomechhamiltonian}), via the position coordinate. The CWL action is then
\begin{widetext}
\begin{equation}
S\;\;=\;\; -i\bar{A}^{j}(t_{f})A_{j}(t_{f})+\int dt\left(i\bar{A}^{j}\dot{A}_{j}-\omega_{\textrm{cav}}\bar{A}^{j}A_{j}+\frac{m}{2}\dot{q}_{j}\dot{q}^{j}-\frac{m\Omega^{2}}{2}q_{j}q_{k}V^{jk}+Gq_{j}\bar{A}^{j}A^{j}\right),
\end{equation}
where we're no longer explicitly writing the classical drive terms.

If we now pass to the CWL (anti-)holomorphic variables, expand about the classical equilibrium, and  perform a variable change to move to a ``co-rotating frame", $a_{j}\rightarrow e^{-i\omega_{\textrm{cav}}t}a_{j}$, $b_{j}\rightarrow e^{-i\Omega t}b_{j}$, we find the CWL action corresponding to the interaction Hamiltonian (\ref{eq:Hintoptomech}) to be
\begin{align}
    S[\{\bar{a}_{j},a_{j},b_{j},\bar{b}_{j}\}] &\;\;=\;\; -i\bar{a}^{j}(t_{f})a_{j}(t_{f})-i\bar{b}^{j}(t_{f})P_{jk}b^{k}(t_{f}) \nonumber \\
    &\qquad  +\int dt\left(i\bar{a}^{j}\dot{a}_{j}+i\bar{b}^{j}P_{jk}\dot{b}^{k}+\frac{\omega_{\textrm{SN}}^{2}}{2\Omega}\frac{\boldone^{j}\boldone^{k}}{N}\bar{b}^{j}b^{k}+g(t)(b_{j}e^{-i\Omega t}+\bar{b}_{j}e^{i\Omega t})(a^{j}e^{i\Delta t}+\bar{a}^{j}e^{-i\Delta t})\right)
\end{align}
with corresponding CWL propagator $K_{fi} \equiv K(\{\bar{\beta}^{(f)}_{j},\bar{\alpha}_{j}^{(f)}\},\{\beta^{(i)}_{j},\alpha_{j}^{(i)}\})$, such that
\begin{align}\label{eq:CWLprop}
    K_{fi}  \;\;=\;\;
    \prod_{j}\left(\int^{e^{-i\Omega t_{f}}\bar{\beta}^{(f)}_j}\mathcal{D}\bar{b}_{j}\int_{\beta^{(i)}_{j}}\mathcal{D}b_{j}
    \int^{e^{-i\omega_{\textrm{cav}} t_{f}}\bar{\alpha}^{(f)}_j}\mathcal{D}\bar{a}_{j}\int_{\alpha^{(i)}_{j}}\mathcal{D}a_{j}\right)\exp\left(iS[\{\bar{a}_{j},a_{j},b_{j},\bar{b}_{j}\}]\right).
\end{align}
\end{widetext}

\subsubsection{Red-detuned pulse}
\label{sec:CWL-Red-pulse}

Let us now consider a finite duration pulse, ie. $g(t)$ which has compact support and laser detuning $\Delta=-\Omega$. If we now apply the standard rotating wave approximation we can retain only the ``beam-splitter interaction terms'', to get
\begin{widetext}
\begin{align}
 \label{eq:optomechaction}
    S[\{\bar{a}_{j},a_{j},b_{j},\bar{b}_{j}\}] & \;\;=\;\; -i\bar{a}^{j}(t_{f})a_{j}(t_{f})-i\bar{b}^{j}(t_{f})P_{jk}b^{k}(t_{f}) \nonumber \\
    & \qquad\qquad\qquad  +\int dt\left(i\bar{a}^{j}\dot{a}_{j}+i\bar{b}^{j}P_{jk}\dot{b}^{k}+\Omega\epsilon^{2}\frac{\boldone^{j}\boldone^{k}}{N}\bar{b}^{j}b^{k}+g(t)(b_{j}\bar{a}^{j}+\bar{b}_{j}a^{j})\right)
\end{align}
\end{widetext}
The form of this action makes the path-integral very simple; we can immediately integrate over the Lagrange multipliers $\bar{a}_{j},\,\bar{b}_{j}$, yielding delta functions enforcing the equations of motion
\begin{align}
    \dot{a}_{j}&\;=\; -ig(t)b_{j} \label{eq:aeom}\\
    P_{jk}\dot{b}^{k}&\;=\; -ig(t)a_{j}-i\Omega\epsilon^{2}\frac{\boldone^{j}\boldone^{k}}{N}b^{k}. \label{eq:beom}
\end{align}

To solve this system of equations, we first define ``mean'' variables $A=N^{-1}\boldone^{j}b_{j}$, $B=N^{-1}\boldone^{j}b_{j}$. We can isolate equations of motion for $A$ and $B$ by summing over $j$ in eqns. (\ref{eq:aeom}), (\ref{eq:beom}), to get
\begin{align}
    \dot{A}&\;=\; -ig(t)B \\
    \dot{B}& \;=\; -ig(t)(1+\epsilon^{2})A-i\Omega\epsilon^{2}B.
\end{align}
The solution is
\begin{widetext}
\begin{equation}
    \begin{pmatrix}
    A(t) \\ B(t)
    \end{pmatrix}
    \;\;=\;\; e^{i\tfrac{\epsilon^{2}}{2}\Omega t}
\begin{pmatrix}
A(0)\cos\Phi(t) -iB(0)\sin\Phi(t)-\tfrac{\epsilon^{2}}{2}(\frac{\Omega t}{\Phi}A(0)-iB(0))\sin\Phi(t) \\ B(0)\cos\Phi(t) -iA(0)\sin\Phi(t)-\tfrac{\epsilon^{2}}{2}(\frac{\Omega t}{\Phi}B(0)+iA(0))\sin\Phi(t)
\end{pmatrix},
\end{equation}
\end{widetext}
where we've defined the integrated pulse strength
\begin{equation}
    \Phi(t) \;\equiv\; (1+\tfrac{\epsilon^{2}}{2})\left|\int_{0}^{t}d\tau g(\tau) \right|,
\end{equation}
and we've assumed that the optomechanical coupling is stronger than the CWL interaction $\Phi(t)\gg\epsilon^{2}\Omega t$. The initial values are the mean values of the initial data for the various replicas,
\begin{align}
    A(0)&\;=\;\frac{1}{N}\sum_{j}\alpha_{j}^{(i)} \nonumber \\
    B(0)&\;=\;\frac{1}{N}\sum_{j}\beta_{j}^{(i)}.
\end{align}
As a sanity check we can see that when $\epsilon=0$ we recover the expected Rabi oscillations between the phonon and photon operators.

Now with the mean variables solved for, we can substitute them back into eqtns. (\ref{eq:aeom}) and (\ref{eq:beom}), to obtain the equations of motion for each replica
\begin{align}
    \dot{a}_{j}&\;=\;-ig(t)b_{j} \label{eq:aeom}\\
    \dot{b}^{j}&\;=\;-ig(t)a_{j}-i\epsilon^{2}\left(\frac{g(t)}{2}A(t)+\Omega B(t)\right). \label{eq:beom}
\end{align}
The solutions are
\begin{widetext}
\begin{align}\label{eq:solutions}
    \begin{pmatrix}
    a_{j}(t) \\ b_{j}(t)
    \end{pmatrix}
    &\;=\;
\begin{pmatrix}
\alpha_{j}^{(i)}\cos\phi(t) -i\beta_{j}^{(i)}\sin\phi(t) \\ \beta_{j}^{(i)}\cos\phi(t) -i\alpha_{j}^{(i)}\sin\phi(t)
\end{pmatrix}
\nonumber \\
& \qquad\qquad\qquad\qquad -i\epsilon^{2}\int^{t}_{0}d\tau\left(\frac{g(\tau)}{2}A(\tau)+\Omega B(\tau)\right)
\begin{pmatrix}
    i\cos\phi(t)\sin\phi(\tau)-i\sin\phi(t)\cos\phi(\tau) \\ \sin\phi(t)\sin\phi(\tau)+\cos\phi(t)\cos\phi(\tau)
    \end{pmatrix}
\end{align}
where $\phi(t)$ is again the uncorrected integrated pulse, ie.,
$\phi(t)=\int^{t}_{0}d\tau g(\tau)$. We can now write the final expression for the CWL propagator in \ref{eq:CWLprop}, viz.,
\begin{equation}\label{eq:finalpropanswer}
    K(\{\bar{\beta}^{(f)}_{j},\bar{\alpha}_{j}^{(f)}\},\{\beta^{(i)}_{j},\alpha_{j}^{(i)}\}) \;\;=\;\; \exp\left(e^{-
    i\omega_{\textrm{cav}}t_{f}}\bar{\alpha}_{j}^{(f)}a^{j}
    (t_{f})+e^{-i\Omega t_{f}}\bar{\beta}_{j}^{(f)}P^{jk}b_{k}
    (t_{f})\right)  \qquad
\end{equation}
\end{widetext}
where $a_{j}(t)$, $b_{j}(t)$ are the solutions given in (\ref{eq:solutions}). This expression can be used to answer a variety of optomechanics questions in CWL theory.

\subsubsection{Fock state probabilities}
 \label{sec:4.C-Fock}

One place we expect to see a signature of CWL theory is when the mechanical oscillator is prepared in a Fock state. The red-detuned regime studied above is perfect for this, as it allows one to effectively transfer quantum states into the oscillator.

Consider the following scenario in standard quantum mechanics. By taking $\epsilon=0$,  eqn. (\ref{eq:solutions}) tells us that excitations will rotate between the mechanical oscillator and the cavity mode.  In particular, if we were to apply a pulse such that $\phi(t_{f})=\pi/2$ we would precisely swap the states of the two oscillators. In principle one could prepare the mechanical oscillator in its ground state, $\beta_{j}^{(i)}=0$, with the cavity in some excited state, $\alpha_{j}^{(i)}\neq0$, and then apply a $\pi/2$ pulse,  effectively loading the cavity state into the mechanical oscillator. This non-trivial state would then evolve coherently, and then another $\pi/2$-pulse could be applied at a later time to transfer the mechanical oscillator state back to the cavity mode. One could then perform a photon count on the cavity mode and if the entire evolution was coherent one expects to recover the initial photon state with high fidelity.

In CWL there are a number of steps in this protocol where the physics may differ from conventional quantum mechanics. Firstly, we've seen that the system dynamics during the pulse are modified such that a $\pi/2$-pulse does not completely swap the oscillator states and this inefficiency would manifest during the both the loading and the readout pulses. Additionally, in the absence of optomechanical coupling there is still a non-trivial modification to the  oscillator dynamics because of the CWL interactions.

Let's actually compute the CWL modifications to this specific process. To be specific, we will again chose the pulse profile to be that given in eqn. (\ref{pulseS}) and shown in Fig. \ref{fig:cavity}(b), which describes a rectangular pulse of duration $t_{p}$, followed by a free evolution for time $T$, and then another rectangular pulse of duration $t_{p}$.

We will assume that pulse sequence duration is short compared with the CWL interaction timescale, ie., so that $2t_{p}+T\ll(\epsilon^{2}\Omega)^{-1}$, and that the mechanical oscillator is initially in its ground state, $\beta^{(i)}_{j}=0$.

With all of these choices the solutions (\ref{eq:solutions}) simplify considerably,
\begin{align}
    a_{j}(2t_p+T)&\;=\;-\alpha_{j}^{(i)}+i\epsilon^{2}\Omega(t_{p}+T)\left(\frac{1}{N}\sum_{j}\alpha_{j}^{(i)}\right) \\
    b_{j}(2t_p+T)&\;=\;i\epsilon^{2}\frac{(2n+1)\pi}{4}\left(\frac{1}{N}\sum_{j}\alpha_{j}^{(i)}\right) \;\;\;\;\;\;
\end{align}

We can see clearly how the CWL corrections alter the result so that the excitation does not completely rotate back into the cavity mode after the pulse sequence. The propagator for the system for this process is thus
\begin{align} \label{eq:CWLfinalpropred}
 K(\{\bar{\beta}^{(f)}_{j}&,\bar{\alpha}_{j}^{(f)}\},\{0,\alpha_{j}^{(i)}\})\;\;=\;\;\exp\left[\zeta^{k}\alpha_{k}^{(i)}\right]
\end{align}
where we've defined
\begin{eqnarray}
  \zeta^{k} &\;&=\; e^{-i\omega_{\textrm{cav}}t_{f}}\bar{\alpha}_{j}^{(f)}\left(-\delta^{jk}+i\epsilon^{2}\Omega(t_{p}+T)\frac{\boldone^{j}\boldone^{k}}{N}\right) \nonumber \\ && \qquad\qquad\qquad  +\; i\epsilon^{2}e^{-i\Omega t_{f}}\bar{\beta}_{j}^{(f)}\frac{(2n+1)\pi}{4}\frac{\boldone^{j}\boldone^{k}}{N} \;\;\;\;\;
\end{eqnarray}

How can we use this result to predict photon counting statistics? It is actually quite simple because the coherent state path-integral is a generating function for Fock amplitudes. This can be seen immediately in operator notation; from an unnormalized coherent state $|\alpha\rangle\equiv e^{\alpha a^{\dagger}}|0\rangle$ one obtains Fock states by simple differentiation
\begin{equation}
    |n\rangle \;\;=\;\; \frac{1}{\sqrt{n!}}\frac{d^{n}}{d\alpha^{n}} |\alpha\rangle\bigg|_{\alpha=0}.
\end{equation}
Fock state transition amplitudes immediately follow
\begin{equation}
    \langle m |\hat{U}(t)|n\rangle \;=\;  \frac{1}{\sqrt{m!n!}}\frac{d^{m}}{d\bar{\beta}^{m}}\frac{d^{n}}{d\alpha^{n}} \langle \beta|\hat{U}|\alpha\rangle\bigg|_{\alpha=\bar{\beta}=0}
\end{equation}

We can utilize this to ``prepare'' a single photon initial state in our optomechanical system by differentiating (\ref{eq:CWLfinalpropred}) with respect to all of the $\alpha_{j}^{(i)}$ and then setting all of the $\alpha_{j}^{(i)}=0$. This is trivial to do, and we obtain the amplitude to evolve from an oscillator in its ground state plus a single cavity photon into CWL coherent states as
\begin{equation}
    K(\{\bar{\beta}^{(f)}_{j},\bar{\alpha}_{j}^{(f)}\},\{0,1\}) \;\;=\; \; \prod_{k=1}^{N}\zeta^{k}.
\end{equation}

To describe a photon counting measurement, we must further differentiate this result. Since the result is a homogeneous polynomial of degree $N$ in the $\bar{\alpha}^{(f)}_{j}$ and $\bar{\beta}_{j}^{(f)}$ there are two replica symmetric possibilities: every replica has one cavity photon and zero mechanical excitations, or zero cavity photons and one mechanical excitation.

The zero photon case is straightforward to compute; one gets
\begin{equation}
    K(\{1,0\},\{0,1\})\;=\;\left[i\epsilon^{2}e^{-i\Omega t_{f}}\frac{(2n+1)\pi}{4}\right]^{N}\frac{N!}{N^{N}}
\end{equation}
The one photon case requires a little more algebra, but the result is
\begin{widetext}
\begin{align}
    K(\{0,1\},\{0,1\})&\;\;=\;\; \frac{d}{d\bar{\alpha}^{(f)}_{1}}...\frac{d}{d\bar{\alpha}^{(f)}_{N}}\left(\prod_{k=1}^{N}\zeta_{k}\right)\bigg|_{\bar{\alpha}^{(f)}_{j}=0} \nonumber \\
    &\;\;=\;\; (-e^{-i\omega_{\textrm{cav}}t_{f}})^{N}\left(1-\frac{ix}{N}\right)^{N}e^{1+i\frac{N}{x}}\left(N-ix\right)^{-N}\Gamma(1+N,1+i\frac{N}{x}),
\end{align}
\end{widetext}
where $x=\epsilon^{2}\Omega(t_{p}+T)$.

We can now take these amplitudes and compute the preprobabilities (ie. the probabilities before we renormalize them to sum to unity). We have
\begin{align}
    p(0)&\;=\;\lim_{N\rightarrow\infty}|K(\{1,0\},\{0,1\})|^{2/N} \nonumber \\
    &\;=\;\; \epsilon^{4}\left(\frac{(2n+1)\pi}{4e}\right)^{2},
\end{align}
and, amusingly,
\begin{align}
    p(1)&\;=\; \lim_{N\rightarrow\infty}|K(\{0,1\},\{0,1\})|^{2/N} \nonumber \\
    &\;=\;\; 1.
\end{align}
The true probabilities are given by a simple renormalization of these
\begin{align}
    P(0)&\;=\; \epsilon^{4}\left(\frac{(2n+1)\pi}{4e}\right)^{2} \nonumber \\
    P(1)&\;=\; 1-P(0)
\end{align}
 Of course, by design, we find probabilities $P(0)=0$ and $P(1)=1$ in the limit of conventional QM. The deviations from QM are the main result of our calculation.

\subsection{Pulsed Optomechanics in Schr\"odinger-Newton}
 \label{PulseO-SN}

\subsubsection{Pulse Protocol Result}
We begin again from the discussion of SN theory given in section 3, where we derived the form for the covariance matrix components under conditions of a weak measurement.  Here we can see that the mass behaves as an oscillator with eigenfrequency $\Omega$.  During the pulsed optomechanics experiment that follows, it continues to act as such an oscillator.   We will adopt the Heisenberg Picture, and use $\hat B$ and $\hat B^\dagger$ as the annihilation and creation operator of this $\omega_q$ effective oscillator, and $\hat A$ and $\hat A^\dagger$ as those of the cavity mode. During the pulse, we have
\begin{align}
\dot {\hat A} & =-i(\omega_m-\Omega) \hat A  - i g_0\sqrt{\frac{\omega_m}{\Omega} }\hat B \label{eqSNA} \\
\dot {\hat B} & = - i g_0\sqrt{\frac{\omega_m}{\Omega} }\hat A  \label{eqSNB}
\end{align}
Here we are in the rotating frame in which $\Omega$ oscillation frequency is removed for both $\hat B$ and $\hat A$, and in this case the operator 
$\hat A $ will have a detuning of $\omega_m -\Omega$.  The $g_0$ here is the designed optomechanical coupling in standard quantum mechanics, such that each pulse will lead to a $(2n+1)\pi$ rotation between the mechanical and optical modes.  It is multiplied by an additional factor $\sqrt{{\omega_m}/{\Omega}} $ since the physical coupling between $Q$ and $\hat A^\dagger \hat A$ is fixed, and now the relation between $\hat Q$ and $\hat B$ is a modified one
\begin{equation}
\hat Q =\sqrt{\frac{\hbar}{m\Omega }}\frac{\hat B +\hat B^\dagger}{\sqrt{2}}
\end{equation} 
and 
\begin{equation}
g  =\lambda \bar A \sqrt{\frac{\hbar}{m\Omega }} 
\end{equation}
while in standard quantum mechanics we have
\begin{equation}
g_0  =\lambda \bar A \sqrt{\frac{\hbar}{m\omega_m}} 
\end{equation}
At time $t$, we have
\begin{align}
\label{SNinout}
& \left[
\begin{array}{c}
\hat A(t)\\
\hat B(t) 
\end{array}\right]\nonumber\\
=&
e^{-\frac{i\Delta t}{2}}
\left[
\begin{array}{cc}
\cos\Phi - \frac{i \Delta \sin\Phi}{\sqrt{4g^2+\Delta^2} } &  -\frac{2ig \sin\Phi}{\sqrt{4g^2+\Delta^2}}\\ 
  -\frac{2ig \sin\Phi}{\sqrt{4g^2+\Delta^2}}  & \cos\Phi + \frac{i \Delta \sin\Phi}{\sqrt{4g^2+\Delta^2} } 
\end{array}\right]
\left[\begin{array}{c}
\hat A(0)\\
\hat B(0)
\end{array}
\right]
\end{align}
with
\begin{equation}
\Phi = \frac{\sqrt{4g^2+\Delta^2} t}{2}\,,
\end{equation}
and
\begin{equation}
g = g_0\sqrt{\omega_m/\Omega},,\quad \Delta = \omega_m-\Omega\,.
\end{equation}
We can invert Eq.~\eqref{SNinout} and take the Hermitian conjugate, obtaining
\begin{align}
\label{SNoutin}
& \left[
\begin{array}{c}
\hat A^\dagger(0)\\
\hat B^\dagger(0) 
\end{array}\right]\nonumber\\
=&
e^{-\frac{i\Delta t}{2}}
\left[
\begin{array}{cc}
\cos\Phi - \frac{i \Delta \sin\Phi}{\sqrt{4g^2+\Delta^2} } &  -\frac{2ig \sin\Phi}{\sqrt{4g^2+\Delta^2}}\\ 
  -\frac{2ig \sin\Phi}{\sqrt{4g^2+\Delta^2}}  & \cos\Phi + \frac{i \Delta \sin\Phi}{\sqrt{4g^2+\Delta^2} } 
\end{array}\right]
\left[\begin{array}{c}
\hat A^\dagger(t )\\
\hat B^\dagger (t)
\end{array}
\right]
\end{align}
Starting from an initial state with only one photon in the cavity, we have
\begin{align}
\hat A^\dagger(0) |0\rangle = e^{-\frac{i\Delta t}{2}} 
\Bigg[\left(\cos\Phi - \frac{i \Delta \sin\Phi}{\sqrt{4g^2+\Delta^2} }\right) \hat A^\dagger(t) \nonumber\\
  -\frac{2ig \sin\Phi}{\sqrt{4g^2+\Delta^2}} \hat B^\dagger(t) 
\Bigg] |0\rangle\,.
\end{align}
where the first term contains one photon in the cavity and the second term contains zero photon in the cavity. 
After two pulses with $(n+1/2)\pi$ rotation, each with
\begin{equation}
t_p =(n+1/2)\pi / g_0
\end{equation} 
 for cavity mode, zero-photon and one-photon probabilities are given by
\begin{equation}
p_0 = \left(\frac{(2n+1) \pi }{4}\right)^2 \epsilon^4\,,\quad 
p_1 = 1- p_0\,.
\end{equation}
Note that here $p_0$ is a factor $e^2$ greater than the CWL case. 

\subsubsection{Finite Temperature Effects}


In the presence of thermal fluctuations, let us consider two effects.  First being the imperfect preparation of the ground state, while the second being thermal disturbances during the pulses.  When considering thermal noise, let us take $\omega_{\rm SN} \rightarrow 0$, and have
\begin{align}
\label{eqAth}
\dot{\hat A } &=- i g_0 \hat B  \\
\label{eqBth}
\dot{\hat B} &= - i g_0 \hat A-\gamma_m \hat B  -\sqrt{2\gamma_m} F_{\rm th}
\end{align} 
Here $F_{\rm th}$ is a classical thermal force acting on the slowly varying amplitude $B$. Assuming that it has a white spectrum (in the case of viscous damping), and imposing that the variance of $B$ caused by $F_{\rm th}$ to be $k_B T/(\hbar\omega_m)$, we obtain a double-sided spectral density of
\begin{equation}
S_{F_{\rm th}} = \frac{2\gamma_m k_B T}{\hbar \omega_m}
\end{equation}
which also leads to 
\begin{equation}
\langle F_{\rm th} (t) F^*_{\rm th}(t')\rangle =   \frac{2\gamma_m k_B T}{\hbar \omega_m}  \delta(t-t')
\end{equation}
Then, solving Eqs.~\eqref{eqAth} and \eqref{eqBth} leads to 
\begin{equation}
\hat A(t) = - \hat A(0) - i \int_0^{2t_p} F_{\rm th}(t')\sin(g t') dt'  \,,
\end{equation}
which leads to 
\begin{equation}
\hat A^\dagger(0) | 0\rangle  = \hat A^\dagger(t) | 0\rangle + i\int_{0}^{2t_p}  F_{\rm th}(t')\sin(g t') dt' |0\rangle  \,.
\end{equation}
This causes a zero-cavity-photon probability of 
\begin{equation}
p_0^{\rm th} =\frac{4\gamma_m t_p k_B T}{\hbar\omega_m } =2\omega_m t_p \frac{k_BT }{\hbar\omega_ m Q}
\end{equation}
The imperfect preparation of ground state for the oscillator gives rise to initial excitations in  $\hat B(0)$ ---  but in the case $\omega_{\rm SN} \rightarrow 0$ that does not show up in the number of cavity photons at $t=2t_p$, because $\hat A(t)$ is unrelated to $\hat B(0)$.  This means imperfect initial-state preparation due to thermal noise is less important than thermal excitations during the pulsed optomechanics process.

In order for the pulsed optomechanics regime to work,  $t_p$ cannot be much less than $1/\omega_m$, hence the requirement for $p_0^{\rm th}$ to be less than unity is 
\begin{equation}
\frac{k_B T}{\hbar\omega_m Q } \stackrel{<}{_\sim}  1\,.
\end{equation}
As we have seen earlier in this section, both CWL and SN theory predicts a $p_0 \sim 1$ when $\omega_{\rm SN}$ is comparable to $\omega_m$. Assuming that thermal noise acts similarly, the will require then conclude that the regime for the CWL and SN experiment to work will be 
\begin{equation}
\omega_m \sim \omega_{\rm SN} \,,\quad \frac{k_B T}{\hbar \omega_{\rm SN} Q} < 1\,.
\end{equation}
For Tungsten, we have  $\omega_{\rm SN} = 2\pi\times 4.0\,{\rm mHz}$, which leads to the following requirements on the mechanical oscillator:
\begin{equation}
\omega_m \sim 2\pi\times 4.0\,{\rm mHz}\,,\quad \frac{T}{Q} <  2\times 10^{-13}\,\textrm{K}\,.
\end{equation} 
In principle this mechanical frequency could be achieved with optomechanical systems based on torsional pendulums. To achieve the necessary temperatures and $Q$-factor however goes beyond present-day capabilities. Given that both SN and CWL both predict departures from QM for this experimental protocol with these experimental parameters, one can certainly view these parameters as an exciting target for future experimental work in quantum optomechanics.

%


\section{Discussion and Conclusions}
 \label{sec:conclude}


The results given in the bulk of this paper are somewhat technical - here we would like to give a more general discussion of what one learns from these calculations. One general conclusion is that it is likely to be quite difficult to test the CWL theory with present day quantum optomechanical technology---however there is a clear target for future experimental parameters which could test both CWL and SN theory.

In section \ref{sec:CWL-factor} we demonstrate the primary obstacle to performing simple tests of CWL, i.e.  for linear measurements and Gaussian states CWL predicts no departures from QM. It follows that:

 (i) to see any observable difference between CWL predictions and ordinary QM predictions for an oscillator, we must use non-Gaussian input states, or else a non-linear measurement of the states must be employed (as in our example of the pulsed optomechanics experiment). Even then, as we have seen, the predicted differences may be very small, because of the large frequency mismatch between typical mechanical oscillator frequencies $\omega$ and the very low Schr\"odinger-Newton frequency $\omega_{\textrm{SN}}$. Nonetheless, in this paper we've performed a concrete computation of the predicted departure from QM for a particular pulse protocol and it unambiguously demonstrates the need for better quantum control over low frequency mechanical oscillators.
 
 (ii) Alternatively, one can employ what are effectively anharmonic effective potentials governing the system dynamics. This opens up a wide set of possibilities which we will not explore here. 

  One might add as a corollary to this that recent observations of `squeezed state' behaviour in a LIGO-type system \cite{mcCuller20} only shows that the mirror dynamics is not classical - they do not yet allow one to discriminate between QM predictions and the predictions of either CWL or SN theory. Under the usual conditions in which LIGO is operated, the behaviour will be indistinguishable from the QM predictions. 

 For SN theory there is not such a strict statement about linear measurements and Gaussian states, and one could aim to test SN using different experimental techniques. It is interesting though that both CWL and SN predict nearly identical probabilities for the pulsed optomechanics experiment discussed herein. This offers the exciting opportunity for a single experiment to simultaneously test both models against QM. 
 
 Regardless of experimental details, it seems clear that the magnitude of the departures from QM will be controlled by the ratio of a characteristic mechanical frequency $\omega$ with the Schr\"odinger-Newton frequency $\Omega_{\rm SN}$, suggesting that these models may be best tested by quantum optomechanics experiments operating in the mHz regime. Our estimates on the required temperature and $Q$-factor are certainly difficult to achieve for such a low frequency oscillator, so it may be that one needs a more innovative experimental protocol than what we've discussed here.

\acknowledgements

J.W.G. is supported by a fellowship at the Walter Burke Institute for Theoretical Physics, a Presidential Postdoctoral Fellowship, the DOE under award number DE-SC0011632, and the Simons Foundation (Award Number 568762). Research of Y.C. is supported by the Simons Foundation (Award No. 568762). PCES was supported in Vancouver by the Natural Sciences and Engineering
Research Council of Canada, No. RGPIN-2019-05582, and at Caltech by U.S.
Department of Energy Basic Energy Sciences Award No. DE-SC0014866.


\end{document}